\documentclass[12pt]{iopart}
\usepackage{subfigure}
\usepackage{graphicx}
\usepackage{multicol}
\usepackage{multirow}
\usepackage{enumerate}
\usepackage{alphalph}
\usepackage{algorithm}
\usepackage{url}
\usepackage{color}
\usepackage[noend]{algpseudocode}

\begin{document}

\title[Leveraging disjoint communities for detecting overlapping community structure]{Leveraging disjoint communities for detecting
overlapping community structure}

\author{Tanmoy Chakraborty}

\address{Department of Computer Science \& Engineering\\ Indian Institute of Technology, Kharagpur, India - 721302}
\ead{its\_tanmoy@cse.iitkgp.ernet.in}
{\color{blue}Accepted at Journal of Statistical Mechanics: Theory and Experiment (JSTAT)}
\vspace{10pt}
\begin{indented}
\item[]April 2015
\end{indented}

\begin{abstract}
Network communities represent mesoscopic structure for understanding the organization of real-world networks, where nodes often belong to
multiple communities and form overlapping community structure in the network. Due to non-triviality in finding the exact boundary of
such overlapping communities, this problem has become challenging, and therefore huge effort has been devoted to detect
overlapping communities from the network.

In this paper, we present {\em PVOC} (Permanence based Vertex-replication algorithm for Overlapping Community detection), a two-stage
framework
to detect overlapping community structure. We build on a novel observation that non-overlapping community structure detected by a standard
disjoint community detection algorithm from a network has high resemblance with its actual overlapping community structure,
except the
overlapping part. Based on this observation, we posit that there is perhaps no need of building yet another overlapping community finding
algorithm; but one
can efficiently manipulate the output of any existing disjoint community finding algorithm to obtain the required overlapping structure. We
propose a new {\em post-processing technique} that by combining with any existing disjoint community detection algorithm, can suitably
process each vertex using a new vertex-based metric, called {\em permanence}, and
thereby finds out overlapping candidates with their community memberships. Experimental results on both
synthetic and
large real-world networks show that PVOC significantly outperforms six state-of-the-art overlapping community detection algorithms in terms
of high similarity of the output with the ground-truth structure. Thus our framework not only finds meaningful overlapping
communities from the network, but also allows us to put an end to the constant effort of building yet another overlapping
community detection
algorithm.

\end{abstract}

%\maketitle

\section{Introduction}
One of the most used aspects of social network analysis  is to discover and display clusters and communities in networks -- the
dense sub-networks, where there are more links internally, than externally. It is easy for the common person to spot dense clusters of
connection in a small network visualization.  However, this is extremely difficult problem to detect such groups from large scale
networks. There has been a constant effort since last one decade from the researchers of both computer science and physics domains to
explore such community
structure from networks after the pioneering effort of Girvan and Newman \cite{NewGir04}. Today there are dozens of
community detection algorithms that can detect the disjoint/non-overlapping community structure from the network using different heuristics
and frameworks (see \cite{Fortunato,Papadopoulos:2012} for the survey). However, in real-world scenario, it has been observed that a node
can be a part of multiple communities, which has eventually led to the idea of overlapping/soft communities
\cite{Raghavan-2007,oslom,Shen,Gregory:2007}. This problem is even more harder because of the exponential number of possible solutions.
Therefore, a new direction of research has been started to detect the overlapping community structure from the network (see \cite{Xie_2013s}
for the survey). 

The dichotomy between ``disjoint'' and ``overlapping'' community detection algorithms is unfortunate because it limits the application of
each algorithm. If a network has overlapping communities, a ``disjoint'' algorithm cannot find them; conversely, if communities are known to
be disjoint, a ``disjoint'' algorithm will generally perform better than an ``overlapping'' algorithm. Therefore, to obtain the actual
community structure, it is important to choose the right kind of algorithm. Note that the question of how to choose the right kind of
algorithm is outside the scope of the present paper.

However, we hypothesize that there is perhaps no need to develop yet another overlapping community finding algorithm given the assumption
that we have diverse and efficient disjoint community detection algorithms in hand. In this paper, we present a method to allow
any ``disjoint'' community detection algorithm to be used to detect overlapping community structure instead for finding another overlapping
community detection algorithm. This means that a user wishing to find overlapping communities need no longer be forced to use one of the
overlapping algorithms that exist, but can also choose from the many disjoint community finding algorithms. The proposed framework is
called as {\em PVOC} (Permanence based Vertex-replication algorithm for Overlapping Community detection) which is a two-phase framework --
in
the first step, an efficient disjoint community detection algorithm is used to detect the non-overlapping community structure from the
network; in the second step, each node in the disjoint communities is processed appropriately using a new vertex-based metric, called
{\em permanence} \cite{chakraborty_kdd}, in order to measure the extent of belongingness of a vertex in its own community and its attached
neighboring communities.  If the membership of the vertex in its assigned community is similar to that in the neighboring community, we
assign the vertex into the neighboring community, keeping its original community intact. Thus the post-processing step is the fundamental
component in PVOC to find out overlapping vertices from the non-overlapping structure. 

We compare our framework with six state-of-the-art overlapping community detection algorithms on both synthetic and large real-world
networks (whose ground-truth community structure is available). We observe that PVOC significantly outperforms other baseline algorithms in
terms of high resemblance of the output with the ground-truth structure. Moreover, we show that even if it is scalable, it does not
compromise the correctness of the output. 

Our paper makes several unique contributions to the state-of-the-art in community detection. These include (i) analyzing the
real-world community structure and observing that the disjoint communities are enough to be processed for discovering overlapping community
structure, (ii) proposing a new framework by combining existing disjoint community detection algorithm along with the post-processing step,
(iii) showing the accuracy of PVOC in terms of accurately discovering the ground-truth structure.

The organization of the paper is as follows. In the next section, we provide a brief overview of state-of-the-art approaches in overlapping
community detection. Section \ref{sec:dataset} provides a brief description of the synthetic and real-world datasets. Following this, in
Section \ref{sec:framework}, we present a detailed results of our empirical observation followed by the description of our proposed
framework. Section \ref{results} describes the results of the experiments to detect overlapping communities and a comparative analysis with
the baseline algorithms. The experiments in this paper use a combination of PVOC with two existing disjoint community detection algorithms,
Louvain \cite{blondel2008} and Infomap \cite{rosvall2007}. Finally, we conclude the paper in Section \ref{conclusion} with some immediate
future directions. 

\section{Related work}\label{related_work}

There has been a class of algorithms for network clustering, which allow nodes belonging to more than one community. Palla proposed
``CFinder'' \cite{PalEtAl05}, the seminal and most popular method based on clique-percolation technique. However, due to the clique
requirement and the sparseness of real networks, the communities discovered by CFinder are usually of low quality \cite{Fortunato:2009}. The
idea of partitioning links instead of nodes to discover community structure has also been explored \cite{nature2010, Evans, evans:2009,
P03021}.

On the other hand, a set of algorithms utilized local expansion and optimization to detect overlapping communities. For instance, Baumes
et al. \cite{BaumesGKMP05} proposed a two-step algorithm ``RankRemoval'' using a local density function. LFM \cite{033015} expands
communities from a random seed node to form a natural community until a fitness function is locally maximal. MONC \cite{abs-1012-1269} uses
the modified fitness function of LFM which allows a single node to be considered a community by itself. OSLOM \cite{oslom} tests the
statistical significance of a cluster with respect to a global null model (i.e., the random graph generated by the configuration model)
during community expansion. Chen et al. \cite{chen2010detecting} proposed selecting a node with maximal node strength based on two
quantities -- belonging degree and the modified modularity. EAGLE \cite{Shen} and GCE \cite{lee} use the agglomerative framework to produce
overlapping communities. COCD \cite{du} first identifies cores and then remaining nodes are attached to cores with which they have maximum
connections. 

Few fuzzy community detection algorithms have been proposed that quantify the strength of association between all pairs of
nodes and communities \cite{Gregory}. Nepusz et al. \cite{Nepusz} modeled the overlapping community detection as a nonlinear constrained
optimization problem which can be solved by simulated annealing methods. Zhang et al. \cite{ZhaWanZha07} proposed an algorithm based on the
spectral clustering framework. Due to the probabilistic nature, mixture models provide an appropriate framework for overlapping community
detection \cite{Newman05062007, citeulike:4205011, citeulike:394155, Zarei}. MOSES \cite{moses} uses a local optimization scheme in which
the fitness function is defined based on the observed condition distribution. Zhang et al. used Nonnegative Matrix Factorization (NMF) to
detect overlapping communities when the number of communities and the feature vectors are provided \cite{PhysRevE.76.046103, 5498458}. Ding
et al. \cite{ding} employed the affinity propagation clustering algorithm for overlapping detection. Recently, BIGCLAM \cite{Leskovec}
algorithm is also built on NMF framework.  

The label propagation algorithm has been extended to overlapping community detection by allowing a node to have multiple labels. In COPRA
\cite{Gregory1}, each node updates its belonging coefficients by averaging the coefficients from all its neighbors at each time step in a
synchronous fashion. SLPA \cite{Xie, abs-1105-3264} spreads labels between nodes according to pairwise interaction rules. A game-theoretic
framework is proposed in Chen et al. \cite{Chen:2010} in which a community is associated with a Nash local equilibrium.

Beside these, CONGA \cite{Gregory:2007} extends GN algorithm \cite{Girvan2002} by allowing a node to split into multiple copies. 
Zhang et al. \cite{ZhangWWZ09} proposed an iterative process that reinforces the network topology and propinquity that is interpreted as the
probability of a pair of nodes belonging to the same community. Istv{\'a}n et al. \cite{pone.0012528} proposed an approach focusing on
centrality-based influence functions. Recently, Gopalan  and Blei \cite{Gopalan03092013} proposed an algorithm that naturally interleaves
subsampling from the network and updating an estimate of its communities. The reader can get more details in a nice survey paper by Xie et
al. \cite{Xie_2013s}.

\section{Test suite of networks}\label{sec:dataset}
\subsection{Synthetic networks}
It is necessary to have good benchmarks to both study the behavior of a proposed community detection algorithm and to compare the
performance across various algorithms. In light of this requirement, Lancichinetti et al. \cite{Lancichinetti} introduced
LFR\footnote{\url{http://sites.google.com/site/andrealancichinetti/files}} benchmark networks that take into account heterogeneity into
degree and community size distributions of a network. These distributions are governed by power laws with exponents $\tau_1$ and $\tau_2$
respectively. To generate overlapping communities $O_n$, the fraction of overlapping nodes is specified and each node is assigned to $O_m$
($\geq$ 1) communities. LFR also provides a rich set of parameters to control the network topology, including the number of nodes $n$, the
mixing parameter $\mu$, the average degree $\bar k$, the maximum degree $k_{max}$, the maximum community size $c_{max}$, and the minimum
community size $c_{min}$. We vary these parameters depending on the experimental needs. Unless otherwise stated, LFR graph is
generated with the following configuration: $\mu$ = 0.2, $N$=10,000, $O_m$=4, $O_n$=5\%; other parameters being set to their default
values. Results shown are the average of 100 runs. 

\begin{figure*}[!h]
\centering{
\scalebox{0.4}{
\includegraphics{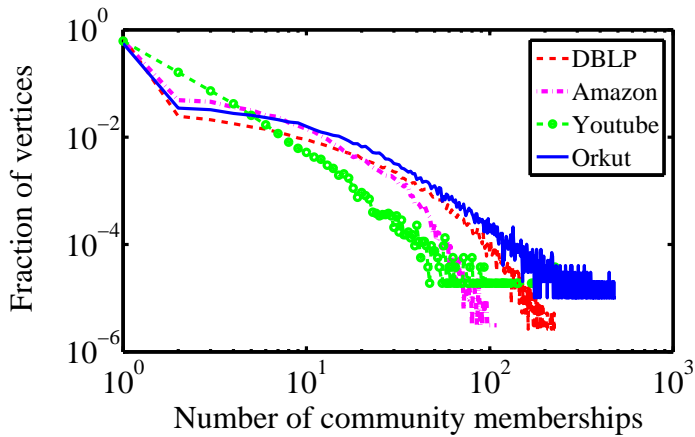}}
\caption{(Color online) Distribution of the number of community memberships of vertices. X-axis shows the number of community memberships of vertices and y-axis shows the fraction of vertices with certain number of community memberships.}\label{dist}}
\end{figure*}

\begin{table}[!h]
\caption{Properties of the real-world networks used in this experiments. $N$: number of nodes, $C$: number of communities, $S$: average
size of a community, $\bar O_m$: average number of community memberships per node.}\label{dataset}
\footnotesize
\centering
 \scalebox{1.1}{
 \begin{tabular}{l||r|r|r|r|r}
 \br
 Networks & N & E & C & S & $\bar O_m$ \\\mr
 DBLP   & 317,080 & 1,049,866 & 13,477  & 429.79 & 2.57 \\\hline
 Amazon & 334,863 & 925,872   & 151,037 & 99.86  & 14.83 \\\hline 
 Youtube & 1,134,890 &	2,987,624 & 8,385 & 9.75 & 10.26 \\\hline
 Orkut  & 3,072,441 & 117,185,083 & 6,288,363 & 34.86 & 95.93\\\hline 
 \end{tabular}}
 
\end{table}

\subsection{Real-world networks with ground-truth communities}
We use four real-world networks\footnote{\url{http://snap.stanford.edu}} proposed by Yang and Leskovec \cite{Leskovec,Yang:2012} whose
underlying ground-truth community structures are known a priori and whose properties are summarized in Table \ref{dataset}.
Figure \ref{dist} shows the distribution of the number of communities memberships of vertices for the real-world networks.
 
{\bf DBLP:} It is a co-authorship network where nodes represent authors and edges connect nodes whose corresponding authors have co-authored
in at least one paper.
Since research communities stem around conferences or journals, the publication venues are used as ground-truth communities in DBLP.

{\bf Amazon:} It is a Amazon product co-purchasing network where nodes represent products and edges connect commonly co-purchased
products. Each product (i.e., node) belongs to one or more product categories. Each product category is used to define a ground-truth
community. 

{\bf Youtube:} In the Youtube social network, users form friendship with each other and users can create groups where other users can join.
Here, such user-defined groups are considered as ground-truth communities.

{\bf Orkut:} Orkut is a free on-line social network where users
form friendship with each other. Orkut also allows users to form a group where other
members can then join. Here also such user-defined groups are considered as ground-truth communities.

\if{0}

\subsection{Scoring functions for evaluating community structure}\label{metric}
The quality of the detected community structure is often measured by certain scoring functions such as modularity, community
coverage, overlap coverage. Below, we provide a brief description of each such function.
\subsubsection{Modularity}
Probably the most widely used measure for evaluating the goodness of a community structure without a
ground truth is Newman's modularity function \cite{NewGir04}. This function is only compatible with disjoint communities in a graph. There
are, however, two variations of modularity designed specifically for evaluating overlapping community structures.

Shen et al. \cite{Shen} introduce $EQ$, an adaptation of Newman's modularity function designed to support overlapping communities. The
equation for $EQ$ strongly resembles the original modularity function as follows:

\begin{equation}
 EQ=\frac{1}{2m}\sum_{c\in C} \sum_{i\in c,j\in c} \frac{1}{O_iO_j} \left[ A_{ij} - \frac{k_ik_j}{2m} \right]
\end{equation}
where $m$ is the number of edges in the graph, $C$ is the set of communities, and $O_v$ is the number of communities to which the node
$v$ belongs. The presence of an edge between two nodes $v$ and $w$ is represented as the value in the corresponding position of the
adjacency matrix $A_{vw}$.

On the other hand, recently L{\'a}z{\'a}r  et al. \cite{Vicsek}  provides a more complex and potentially more accurate evaluation of the
goodness of an overlapping community structure as follows:

% \begin{equation}
%  Q_{ov}=\frac{1}{K} \sum\limits_{r=1}^{K} \Bigg [  \frac{\sum\limits_{i\in c_r}  \frac{\sum\limits_{j\in c_r, i \neq j} A_{ij} -
% \sum\limits_{j \notin c_r} A_{ij}}{di\cdot s_i } }{n_{c_r}} \cdot  \frac{n^e_{c_r}}{	\dbinom{n_{c_r}}{2}}  \Bigg ]
% \end{equation}

where $K$ is the number of communities, $n_{c_r}$ is the number of nodes and $n^e_{c_r}$ is the number of edges that the $r$th cluster $c_r$
contains respectively, $d_i$ is the degree of node $i$, $s_i$ denotes the number of clusters where $i$ belongs to and $A$ is the adjacency
% matrix. Note that, since the density of clusters containing one single node (when $n_{c_r}$ = 1) is not defined (because $\dbinom{1}{2}$ is
undefined), we simply set their modularity value to be zero.

\subsubsection{Community Coverage (CC)} As described in \cite{nature2010}, it simply counts the fraction of nodes that belong to at least
one community of three or more nodes. A size of three is chosen since it is the smallest nontrivial community. This measure provides a
sense of how much of the network is analyzed. 

\subsubsection{Overlap Coverage (OC)} As described in \cite{nature2010}, it counts the average number of memberships in nontrivial
communities (size at leas three) that nodes are given.

\subsection{Community detection algorithms}\label{algo}
We use the following state-of-the-art overlapping community detection algorithms to compare our proposed algorithm:
\begin{itemize}
 \item Order statistics local optimization method (OSLOM)\footnote{\url{http://www.oslom.org.}} \cite{oslom}
 \item Community overlap propagation algorithm (COPRA)\footnote{\url{http://www.cs.bris.ac.uk/~steve/networks/software/copra.html.}}
\cite{Gregory1}
 \item Speaker listener propagation algorithm (SLPA)\footnote{\url{https://sites.google.com/site/communitydetectionslpa.}} \cite{Xie}
 \item Model-based overlapping seed expansion algorithm (MOSES)\footnote{\url{http://sites.google.com/site/aaronmcdaid/moses.}}
\cite{moses}
 \item Agglomerative hierarchical clustering based on maximal clique (EAGLE)\footnote{\url{http://code.google.com/p/eaglepp/}}
\cite{Shen}
 \item Cluster Affiliation Model for Big Networks (BIGCLAM)\footnote{\url{http://snap.stanford.edu}} \cite{Leskovec}
\end{itemize}
Note that, each algorithm is simply used with its default parameters. 

\subsection{Metrics to compare with ground-truth}\label{validation_metrics}
A stronger test of the correctness of the community detection algorithm, however, is by comparing the obtained community with a given
ground-truth structure. For evaluation, we use three metrics that quantify the level of correspondence between the detected and the ground-
truth communities \cite{Leskovec}. 

\begin{itemize}
 \item Overlapping Normalized Mutual Information \cite{McDaid}
 \item Omega Index \cite{Gregory}
 \item Average F1 score \cite{manning}
\end{itemize}

Note that, all the metrics are bounded between 0 (no matching) and 1 (perfect matching).

\fi

\section{Vertex-replication algorithm }\label{sec:framework}
Our proposed algorithm is motivated from an empirical study on the ground-truth community structure of both synthetic and real-world
networks. In this section, we first describe the empirical observation and then illustrate a new
algorithm that can detect overlapping communities from a network with the help of {\em any} standard disjoint community detection
algorithm.

\begin{figure*}[!t]
\centering{
\scalebox{0.35}{
\includegraphics{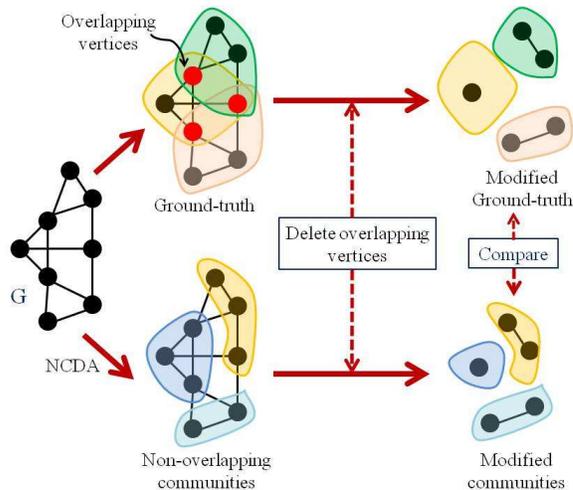}}
\caption{(Color online) An illustrative example to show the procedure followed in our empirical study (NCDA: Non-overlapping Community
Detection Algorithm). }\label{demo}}
\end{figure*}

\subsection{Empirical observation} We empirically study the structure of the ground-truth communities. We speculated that
if we remove the vertices that are part of multiple communities from the ground-truth structure, the rest of the portion, i.e., the
community structure composed of only non-overlapping vertices can be efficiently captured by the standard disjoint community
detection algorithm. To verify this intuition, we take
all the  networks with their ground-truth communities and two standard disjoint community detection algorithms, namely
Louvain\cite{blondel2008} and Infomap\cite{rosvall2007,Rosvall29012008}. Then for each network, we run the following steps:

\begin{enumerate}[I]
 \item We run each of these algorithms to obtain the disjoint community structure from the network.
\item Since we know the ground-truth community structure of the network, we remove from the ground-truth  those vertices (refer to set
$V_o$) which belong to multiple communities.
\item Similarly, we remove the constituent vertices of $V_o$ from the disjoint community structure obtained from Step I. This step
makes sure that the filtered ground-truth community structure and the filtered disjoint community structure obtained from the
algorithm contain same set of vertices.
\item Then we compare two community structures obtained from Step II and Step III. 
\end{enumerate}

\begin{table}
 \centering
\caption{Number of communities in the ground-truth structure and that obtained from Louvain and Infomap for LFR and real-world
networks.
Here for the LFR network, we consider the following configuration: $N$=10,000, $\mu$=0.2, $O_m$=4, $O_n$=5\%. The result of LFR is averaged
over 100 runs.}\label{stat}
\scalebox{0.8}{
\begin{tabular}{|c|c|c|c|}
\hline
  Networks & Ground-truth & \multicolumn{2}{c|}{Algorithms} \\\cline{3-4}
           &               &   Louvain &   Infomap \\\hline
LFR        & 582 & 468 & 501 \\\hline
DBLP & 8,493 & 7,987 & 8,145 \\\hline
Amazon & 151,037 & 142,098 & 149,876\\\hline
Youtube & 8,385 & 7,967 & 7,132 \\\hline
Orkut & 288,363 & 284,980 & 286,791 \\\hline
        
\end{tabular}}
\end{table}

\begin{figure*}[!h]
\centering{
\scalebox{0.21}{
\includegraphics{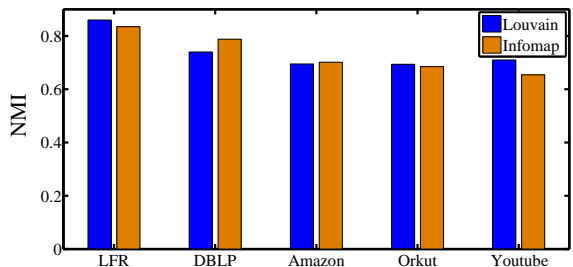}}
\caption{(Color online) Similarity (in terms of NMI) of the community structure obtained from two disjoint community detection
algorithms (Louvain and Infomap) with the ground-truth structures after excluding the overlapping vertices. Here for the LFR network, we
consider the following configuration: $N$=10,000, $\mu$=0.2, $O_m$=4, $O_n$=5\%.}\label{nmi}}
\end{figure*}

A schematic example of the above procedure is shown in Figure \ref{demo}.  Figure \ref{dist} shows that in this process, we
discard nearly 40\% of the vertices (on an average) which belong to multiple communities for each network. In Table \ref{stat}, we also
report the number of disjoint communities obtained from Louvain and Infomap algorithms for both synthetic and real-world networks and that
present in the ground-truth structure. We use a standard validation metric,
namely Normalized Mutual Information (NMI) \cite{danon2005ccs}  to
compare these two community structures. Figure \ref{nmi} shows that the similarity is quite high for all the networks; this observation
indeed
corroborates
our earlier speculation. Therefore, we hypothesize that a standard disjoint community detection algorithm might be able to find the
overlapping
communities with a suitable post-processing step. This means that a user wishing to find overlapping communities need no longer be forced to
use any overlapping community
finding algorithm, rather a disjoint community structure followed by a {\em post-processing step} might produce the expected
overlapping community structure. In the rest of this section, we shall use this observation to design a suitable post-processing technique.

\subsection{Permanence based vertex-replication algorithm}
Through careful inspection mentioned above, we have found that a standard disjoint community
detection algorithms are quite efficient to detect the non-overlapping part of the community structure.
However there exist few vertices in the network, which are part of multiple communities. We intend to design an efficient algorithm that
would
be able to identify such overlapping vertices with their community memberships. For that, we use a vertex-based metric, called {\em
permanence}, which by virtue of its underlying formulation measures how intensely a vertex belongs to its community
\cite{chakraborty_kdd}. Below, we present a brief overview of the formulation of permanence.

\begin{figure}[!h]
\centering
 \includegraphics[scale=0.4]{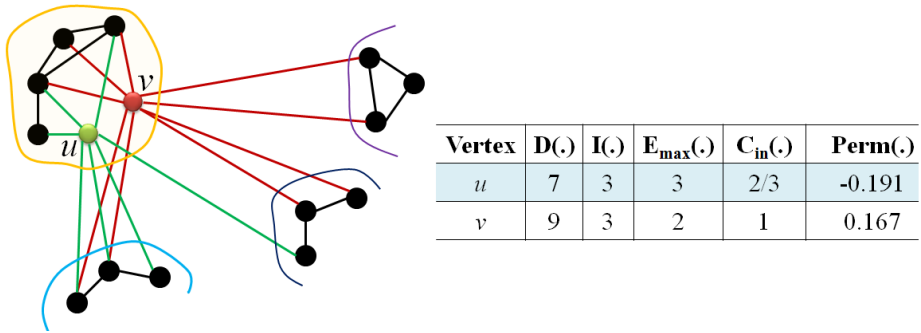}
 \caption{Toy example depicting {\em permanence} of two vertices $u$ and $v$. Even if vertex $v$ has  a large number of
external
connections than $u$, all these six external connections are distributed equally into three neighboring communities,
resulting in the external pull proportional to 2; whereas $u$ is attached with 4 external neighbors, three of them constitute in
one community and the rest is attached with another community, resulting in the external pull proportional to 3. On the other hand, $v$ is
connected to 3 internal neighbors which are further completely connected among each other; whereas the neighbors of $u$ are partially
connected. This results in high internal pull of $v$ as compared to $u$. These two notions of connectively are considered in
the formulation of permanence.}\label{perm_example}
\end{figure}

\subsubsection{Formulation of permanence}
In an earlier paper \cite{chakraborty_kdd}, we showed that the extent of membership of a vertex to a community depends on the following
two factors. (i) The first factor is  {\em the distribution of external connections of the vertex to individual communities}. A
vertex that has equal number of connections to all its external communities (e.g., a vertex with total 6 external connections with 2 to each
of 3 neighboring communities) has equal ``pull" from each community whereas a vertex with more external connections to one particular
community  (e.g., a vertex with total 6 external connections with 1 connection each to two neighboring communities and 4 connections to the
third neighboring community), will experience more ``pull" from that community due to large number of external connections to it. (ii) The
second factor is {\em the density of its internal connections}. The internal connections of a community are generally considered together as
a whole. However, how strongly a vertex is connected to its internal neighbors can differ. To measure this internal connectedness of a
vertex, one can compute the clustering coefficient of the vertex with respect to its internal neighbors. The higher this internal clustering
coefficient, the more tightly the vertex is connected to its community.

 Combining these two factors together, we formulated permanence $Perm(.)$ of a vertex $v$ as follows: 
\begin{equation}\label{perm}
Perm(v)= \frac{I(v)}{E_{max}(v)} \times  \frac{1}{D(v)} - (1-c_{in}(v))
\end{equation}

where $I(v)$ is the number of internal connections of $v$, $D(v)$ is the degree of $v$, $E_{max}(v)$ is the maximum connections of $v$ to a
single external community and $c_{in}(v)$ is the clustering coefficient among the internal neighbors of $v$. An illustrative example is
shown in Figure \ref{perm_example}.

For vertices that do not have any external connections, $Perm(v)$ is considered to be equal to the internal clustering coefficient (i.e.,
$Perm(v) = c_{in}(v)$). The maximum value of $Perm(v)$ is 1 and is obtained when vertex $v$ is an internal node and part of a clique. The
lower bound of $Perm(v)$ is close to -1. This is obtained when $I(v) \ll D(v)$, such that $\frac{I(v)}{D(v)E_{max}(v)} \approx 0$ and  $
c_{in}(v)=0$. Therefore for every vertex $v$, $-1 < Perm(v) \leq 1$.

\begin{algorithm}[t]
\caption{PVOC: Permanence based vertex-replication algorithm for overlapping community detection}\label{algo}
{\bf Input:}  A graph $G=(V,E)$; $A_d$= disjoint community detection algorithm; $\theta=$threshold

{\bf Output:} Detected overlapping communities\\

 \begin{algorithmic}
   \Procedure{Vertex\_replication}{$G$, $NC$, $\theta$}\label{d}
   \State $V_e=$ set of vertices having at least one external neighbor
   \ForAll $\ v \in V_e$
     \State $C_v=$ current community of $v$
     \State Measure current permanence of $v$, $O_p(v)$
     \State Measure current sum of permanences of all neighbor's of $v$, $O_n(v)$
     \State $Sum\_{O_p}=O_p(v)+O_n(v)$
     \ForAll $\ n \in N$ \Comment {N is the set of external neighbors of $v$}
	\State $C_n=$current community of $n$
	\State Remove $v$ from $C_v$ and assign it in $C_n$ 
	\State Measure new permanence of $v$ in community $C_n$, $N_p(v)$
	\State Measure new sum of permanences of all neighbor's of $v$, $N_n(v)$
        \State $Sum\_{N_p}= N_p(v) + N_n(v)$
	\If {($|Sum\_{N_p} - Sum\_{O_p}|<=\theta$)}
	    \State Assign a replica of $v$ in $C_n$ along with its original presence in $C_v$ 
	\Else
	    \State Remove vertex $v$ from $C_n$ and place it back in $C_v$
	 \EndIf
	  
     \EndFor
   \EndFor
\Return The updated overlapping community structure
 
 \EndProcedure

   \Procedure{PVOC}{}
          \State Run $A_d$ on $G$ to obtain disjoint community structure $NC$ 
          \State Call VERTEX\_REPLICATION($G$,$NC$,$\theta$)

    \EndProcedure
 \end{algorithmic}
 \end{algorithm}

\subsubsection{The PVOC algorithm}
Since permanence can assign a score to each of the vertices, we can use it in our post-processing step to identify overlapping vertices
from the detected disjoint community structure. Subsequently, we develop a new algorithm, called {\em PVOC} ({\bf P}ermanence based {\bf
V}ertex-replication algorithm for {\bf O}verlapping {\bf C}ommunity detection) that can combine any existing disjoint community detection
algorithm with the permanence based vertex-replication for detecting overlapping community structure of a network. Algorithm \ref{algo}
presents the pseudo-code of PVOC.  

Given undirected network $G (V, E)$ and a threshold $\theta$, the algorithm works as follows:
\begin{enumerate}[I]
 \item A standard disjoint community detection algorithm $A_d$ is used to detect non-overlapping community structure $NC$ from $G$.
 \item A set of vertices $V_e$ are identified from $NC$ such that each constituent vertex in $V_e$ has at least one connection to
any external community.
  \item For each vertex $v$ in $V_e$, we do the following steps:
     \begin{enumerate}[(a)]
      \item We calculate the sum of permanence of $v$ and its neighbors in their assigned communities.
      \item We remove $v$ from its own community and place it to each of its external communities separately. This assignment affects the
permanence value of $v$ and its immediate neighbors. 
      \item For each external community $C_n$, we measure the current sum of permanence of $v$ (in its new community) and its neighbors.
      \item If the absolute value of the difference of the permanence values obtained from Step III(a) and Step III(c) is less than
$\theta$, a replica of $v$ is placed into the new community $C_n$, keeping $v$ in its original community as well; otherwise $v$ is assigned
back to its original community. This step
identifies overlapping nodes along with their memberships in different communities.
     \item The algorithm finally returns all the vertices with new community membership. 
     \end{enumerate} 
\end{enumerate}

\begin{figure*}[!h]
\centering{
\scalebox{0.4}{
\includegraphics{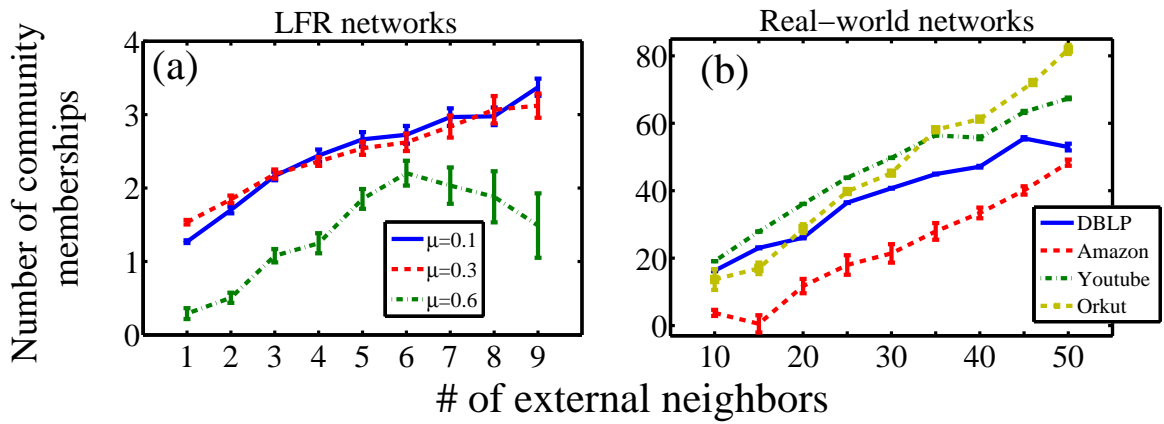}}
\caption{(Color online) The relation between the average number of external connections of vertices (with the standard deviation) obtained
from the output of the disjoint community detection algorithm (here we use Louvian) and the number of communities a vertex is a part of. 
}\label{reation}}
\end{figure*}

The threshold $\theta$ controls the extent to which one can relax the condition of replicating a vertex into multiple communities. We vary
the threshold from 0 to 0.2 and observe that it produces maximum accuracy at 0.05 (see Figure \ref{theta}). Therefore, for the rest of the
experiment, we keep the value of $\theta$ as 0.05.  Note that in the permanence-based post-processing step, we only consider those vertices
having at least one external connection. The rationale behind this assumption is that vertices in the core of each
community are often considered to be correctly placed by the disjoint community detection algorithm, whereas vertices which are placed in
the
peripheral region
of the community and are loosely connected to the core of
the community have high chance to be part of multiple communities. Figure \ref{reation} shows an empirical observation where we
plot the
relation
between the number of external connections of a vertex in the detected disjoint community to the number of overlapping memberships of the
vertex in ground-truth community. We observe that the correlation is increasing in nature, which indeed strengthens our hypothesis.  

The time complexity of measuring the permanence of a vertex takes $O(d^2)$, where $d$ is the average degree of vertices in the network. In
real-world networks, the value of $d$ is much lower than $log(n)$, where $n$ is the number of nodes in the network. Therefore, the PVOC
algorithm mostly depends on the underlying disjoint community detection algorithm.

\section{Experiments}\label{results}
We combine PVOC with two popular disjoint community detection algorithms, namely
Louvain\footnote{\url{https://sites.google.com/site/findcommunities/}} \cite{blondel2008} and
Infomap\footnote{\url{http://www.tp.umu.se/~rosvall/code.html}} \cite{rosvall2007,Rosvall29012008}. These are chosen because they are
reasonably accurate algorithms with the potential to handle large networks, and implementation of them, by their authors, are publicly
available.

\subsection{Baseline algorithms}
We compare the performance of PVOC with the following state-of-the-art overlapping community detection algorithms, whose codes are
also available:
 \begin{itemize}
 \item Order statistics local optimization method (OSLOM): It is based on the local optimization of
a fitness function expressing the statistical significance of clusters with respect to random fluctuations, which is estimated with tools of
Extreme and Order Statistics \cite{oslom}. The code is available at \url{http://www.oslom.org.}

 \item Community overlap propagation algorithm (COPRA): This
algorithm  is based on the label propagation
technique of Raghavan et al \cite{Raghavan-2007}, but is able to detect communities that overlap. Like the original algorithm, vertices have
labels that propagate between neighboring vertices so that members of a community reach a consensus on their community membership
\cite{Gregory1}. The code is available at \url{http://www.cs.bris.ac.uk/~steve/networks/software/copra.html}.

 \item Speaker listener propagation algorithm (SLPA): The algorithm is an extension of the Label Propagation Algorithm (LPA)
\cite{Raghavan-2007}. In SLPA, each node can be a listener or a speaker. The roles are switched depending on whether a node serves as an
information provider or information consumer. Typically, a node can hold as many labels as it likes, depending on what it has experienced in
the stochastic processes driven by the underlying network structure. A node accumulates
knowledge of repeatedly observed labels instead of erasing all but one of them. Moreover, the more a node observes a label, the more likely
it will spread this label to other nodes \cite{Xie}. The code is available at \url{https://sites.google.com/site/communitydetectionslpa}.

 \item Agglomerative hierarchical clustering based on maximal clique (EAGLE): It uses the
agglomerative framework to produce a dendrogram. First, all maximal cliques are found and made to be the initial communities. Then, the pair
of communities with maximum similarity is merged. The optimal cut on the dendrogram is determined by the extended modularity with a weight
based on the number of overlapping memberships \cite{Shen}. The code is available at \url{http://code.google.com/p/eaglepp/}.

 \item Cluster-overlap Newman Griven algorithm (CONGA): The idea of
this algorithm is similar to our idea of finding overlapping
communities from disjoint community structure. CONGA is based on Griven Newman's ``GN'' algorithm \cite{NewGir04} but extended to detect
overlapping communities. CONGA adds to the GN algorithm the ability to split vertices between communities, based on the new concept of
{\em split betweenness}. At first,  edge betweenness of edges and split betweenness of vertices are calculated. Then an edge with
maximum edge betweenness is removed or a vertex with maximum split betweenness is split. After this step, edge betweenness and split
betweenness are recalculated. The above steps are repeated until no edges remain \cite{Gregory:2007}. However, the calculation of edge
betweenness and split betweenness is expensive on large networks. The code is available at
\url{http://www.cs.bris.ac.uk/~steve/networks/congapaper/}.

 \item Cluster affiliation model for big networks (BIGCLAM): In this algorithm, communities
arise due to shared community affiliations of nodes. Here the affiliation strength is explicitly modeled for each node to each community.
Then each node-community pair is assigned a nonnegative latent factor which represents the degree of membership of a node to the community.
The probability of an edge between a pair of nodes is then modeled in the network as a function of the shared community affiliations
\cite{Leskovec}. The code is available at \url{http://snap.stanford.edu}.
\end{itemize}
Note that each algorithm is simply used with its default parameters. 

\begin{figure*}[!h]
\centering{
\scalebox{0.4}{
\includegraphics{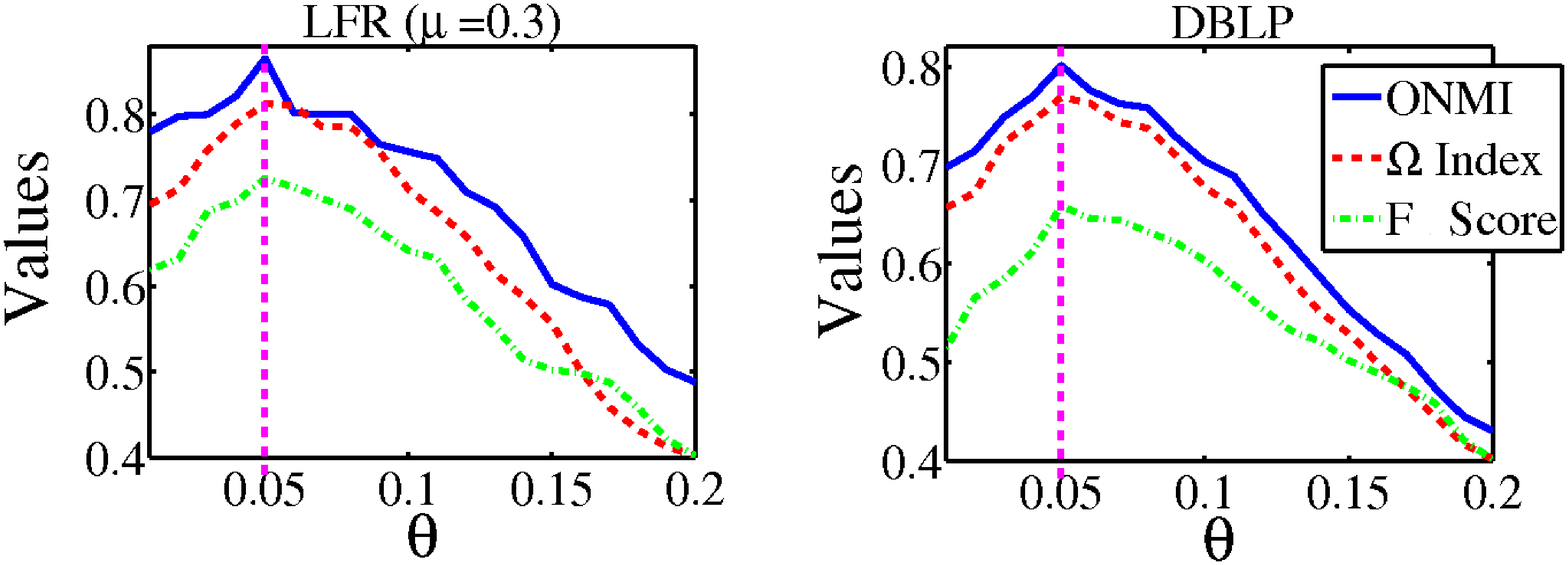}}
\caption{(Color online) Accuracy of PVOC (in terms of three validation metrics) with the increase of $\theta$ for LFR ($\mu=0.3$) and one
real-world network (DBLP). Maximum accuracy is obtained at $\theta=0.05$, which we use in rest of the experiment. Each point in the plot is
 an average of the accuracies obtained from Louvain and Infomap.}\label{theta}}
\end{figure*}

\subsection{Validation metrics}\label{metrics}
A stronger test of the correctness of the community detection algorithm, however, is by comparing the obtained community with a given
ground-truth structure. For evaluation, we use three metrics that quantify the level of correspondence between the detected and the ground-
truth communities \cite{Leskovec}. 

\begin{itemize}
 \item Overlapping Normalized Mutual Information\footnote{\url{https://github.com/aaronmcdaid/Overlapping-NMI}} \cite{McDaid}
 \item Omega Index \cite{Gregory}
 \item Average F score \cite{manning}
\end{itemize}

Note that all the metrics are bounded between 0 (no matching) and 1 (perfect matching).

 \begin{figure*}[!t]
\centering
\scalebox{0.4}{
\includegraphics{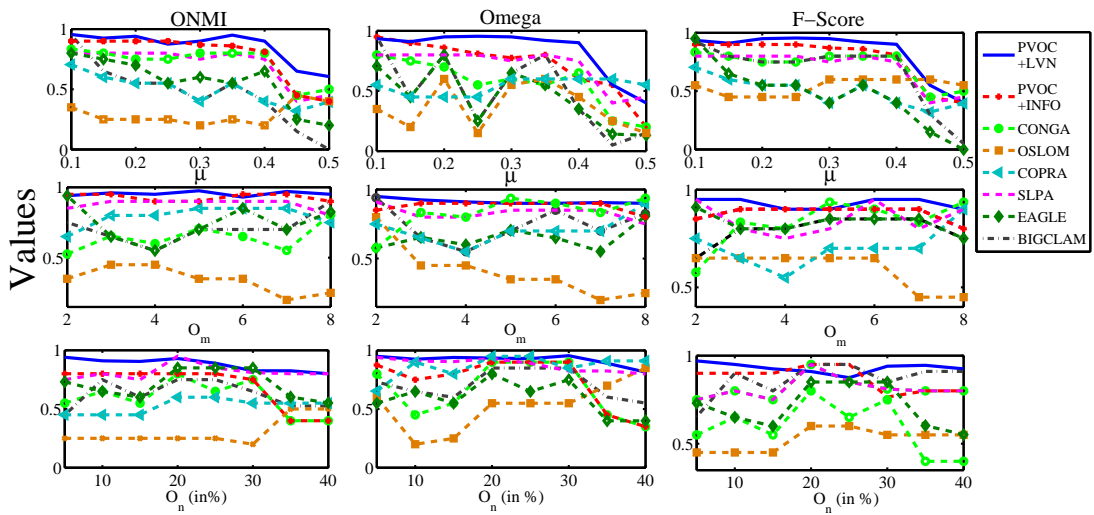}}
\caption{(Color online) Accuracy of all the competing algorithms for LFR by varying $\mu$ (top panel, where $N =10,000$, $O_m =4$, $O_n
=5\%$), $O_m$
(middle point, where $N =10,000$, $\mu=0.1$, $O_n =5\%$) and $O_n$ (bottom panel, where $N =10,000$, $\mu=0.1$, $O_m =4$)). Note that the
value of $O_n$ is expressed
in \% of $n$ (LVN: Louvain, INFO: Infomap).
}\label{lfr}
\end{figure*}

\subsection{Experimental results}
In this experiment, we use PVOC combined with Louvain and Informap separately, and compare the results with six baseline algorithms. 
First, we check the dependency of PVOC with the value of $\theta$. Figure \ref{theta} shows that at $\theta=0.05$, PVOC achieves maximum
accuracy for LFR and one representative real-world network; however the result is almost same of other networks. Therefore, we use
$\theta=0.05$ in the rest of the experiments. One can tune $\theta$ appropriately to control the extent of overlapping
membership of vertices in the network. 

In Figure \ref{lfr}, we compare the outputs obtained from different competing algorithms with the ground-truth communities for
LFR networks with different parameter settings.  Figure \ref{lfr} (top panel) shows the
results for different values of $\mu$ ranging from 0.1 to 0.5. As $\mu$ increases, the community structure becomes less evident and it
becomes
difficult for all the algorithms to discover the actual community structure. OSLOM performs worst compared to the other algorithms. However,
for all the cases, PVOC+LVN is least affected and outperforms other algorithms. This is followed by PVOC+INFO, CONGO and SLPA.

We then vary the average number of community memberships per vertex, $O_m$ from 2 to 8 keeping the other parameters same, and plot the
performance of different algorithms in Figure \ref{lfr} (middle panel). The effect is reasonably less on the accuracy of the competing
algorithms. Here we observe that the pattern is almost similar for PVOC+LVN and PVOC+INFO, and are much superior than others.

Finally, in Figure \ref{lfr} (lower panel) we plot the accuracy of the algorithms with the increasing value of $O_n$, percentage of
overlapping vertices. Surprisingly, OSLOM shows an unexpected behavior with the increasing accuracy after a certain value of $O_n$.
However, on an average the change in accuracy is almost consistent for all the algorithms in all possibilities of $O_n$.

To understand the utility of including PVOC step with the disjoint community finding algorithms in more details, we further
measure the performance of Louvain and Infomap in isolation without PVOC step. We observe that excluding PVOC step significantly
deteriorates the performance of Louvain algorithm: for LFR network ($N$ = 10,000, $O_m$ = 4, $O_n$ = 5\% , $\mu$=0.2) ONMI (0.569), Omega
Index (0.512), F-Score (0.523); for DBLP network ONMI (0.495), Omega Index (0.521), F-Score (0.487); for Amazon network ONMI (0.458), Omega
Index (0.498), F-Score (0.447); for Youtube network ONMI (0.512), Omega Index (0.522), F-Score (0.564); and for Orkut network ONMI (0.526),
Omega Index (0.556), F-Score (0.544). Similar trend is observed for Infomap algorithm. This observation therefore strengthens the need of
PVOC as a post-processing step with the disjoint community detection algorithms.

Now, we run the competing algorithms on the real-world networks. As noted in  \cite{Leskovec}, most of the baseline community detection
algorithms do not scale for networks of large size. Therefore, we use the following technique proposed by Yan and Leskovec \cite{Leskovec}
to obtain several small subnetworks with overlapping community structure from the large real networks. We pick a random node $u$ in the
given graph $G$ that belongs to at least two communities. We then take the subnetwork to be the induced subgraph of $G$ consisting of all
the nodes that share at least one ground-truth community membership with $u$. In our experiments, we created 500 different subnetworks for
each of the six real-world datasets and the results are averaged over these 500 samples. For each validation metric (ONMI,  $\Omega$ Index,
F-Score), we separately scale the scores of the methods so that the best performing community detection method has the score of $1$.
Finally, we compute the composite performance by summing up three normalized scores. If a method outperforms all the other methods in all
the scores, then its composite performance is $3$.

\begin{figure*}[!t]
\centering
\scalebox{0.4}{
\includegraphics{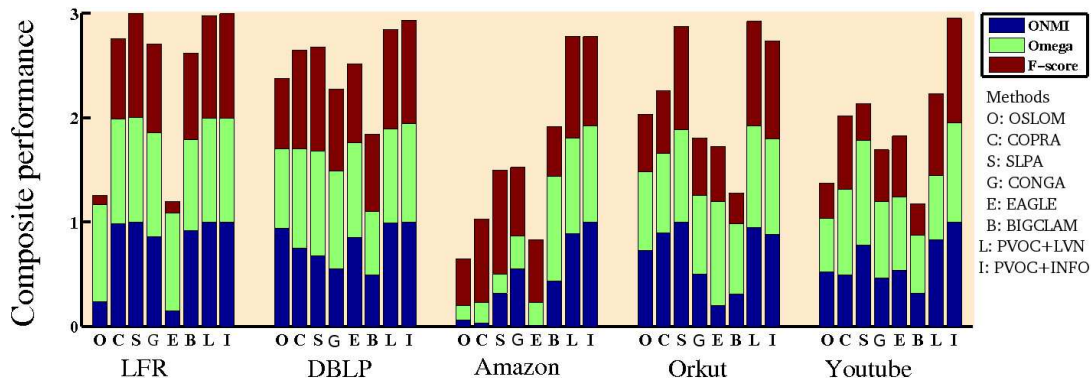}}
\caption{(Color online) Performance of various competing algorithms to detect the ground-truth communities. For each evaluation metric
separately we scale the score of the methods so that the best performing community detection algorithm achieves the score of 1. Thus if an
algorithm outperforms all the methods in all the scores, then its composite score would become 3.}\label{composite_perf}
\end{figure*}

Figure \ref{composite_perf} displays the composite performance of the methods for different networks. On an average, the composite
performance
of PVOC+INFO (2.88) and PVOC+LVN (2.74) significantly outperform other competing algorithms: 6.27\% higher than that of BIGCLAM (2.71),
18.03\% higher than that
of SLPA (2.44), 101.3\% higher than that of OSLOM (1.43), 36.4\% higher than that of COPRA (2.11), 48.4\% higher than that of CONGA (1.94),
and 77.8\% higher than that of EAGLE (1.62). The absolute average ONMI of PVOC+INFO (PVOC+LVN) for one LFR and six real networks taken
together is 0.85 (0.83),
which is
4.93\% (2.46\%) and 26.8\% (20.8\%) higher than the two most competing algorithms, i.e., BIGCLAM (0.81), and SLPA (0.67)
respectively. In terms of absolute values of
scores, PVOC+INFO (PVOC+LVN) achieves the average F-Score of 0.84 (0.79) and average $\Omega$ Index of 0.83 (0.82). Overall, PVOC combined
with Louvain and Infomap gives the best results, 
followed by BIGCLAM, SLPA, COPRA, CONGO, EAGLE and OSLOM.

 As most of the baseline algorithms except BIGCLAM do not scale for large real networks \cite{Leskovec}, we separately compare PVOC
with 
BIGCLAM (which is scalable and also the most competing algorithm) on actual large real datasets. Table \ref{imp}
shows performance of PVOC and BIGCLAM for different real networks. On average, PVOC+INFO (PVOC+LVN) achieves 4.28\% (5.63\%) higher ONMI,
1.48\% (2.85\%) higher $\Omega$ Index, and 6.94\% (5.63\%) higher F-Score. Overall, PVOC outperforms BIGCLAM in every measure and for
every network. The absolute values
of the scores of PVOC+INFO and PVOC+LVN averaged over all
the networks are 0.70 and 0.71 (ONMI), 0.69 and 0.70 ($\Omega$ Index), and 0.72 and 0.71 (F-Score) respectively.

\begin{table}[!h]
\centering
\caption{The performance of BIGCLAM and PVOC on large real-world networks.}\label{imp}
\scalebox{0.7}{
 \begin{tabular}{|c|c|c|c|c|c|c|c|c|c|}
\hline
\multirow{2}{*}{Networks} & \multicolumn{3}{c|}{BIGCLAM} & \multicolumn{3}{c|}{PVOC+LVN} & \multicolumn{3}{c|}{PVOC+INFO}\\\cline{2-10}
               & ONMI & Omega & F Score & ONMI & Omega & F Score & ONMI & Omega & F Score \\\hline   
DBLP           & 0.61 & 0.59  & 0.54    & {\bf 0.65} & 0.61  & {\bf 0.60}    & {\bf 0.65} & {\bf 0.62}  & 0.59   \\\hline
Amazon         & {\bf 0.73} & 0.69  & 0.74    & 0.72 & 0.71  & 0.75    & {\bf 0.73} & {\bf 0.74} & {\bf 0.76}\\\hline
Orkut          & 0.65 & 0.68  & 0.64    & 0.72  & 0.70 & 0.76    & {\bf 0.73} & {\bf 0.72} & {\bf 0.77} \\\hline
Youtube        & 0.68 & 0.76  & {\bf 0.78}    & {\bf 0.77} & {\bf 0.78}  & 0.72    & 0.71 & 0.68 & {\bf 0.78} \\\hline

 \end{tabular}}

\end{table}

Many optimization algorithms have the tendency to underestimate smaller size communities \cite{Barthelemy} and sometimes tend to produce
very large size communities. In our test suite, we observe the similar tendency in BIGCLAM whereas the communities
obtained by PVOC based algorithms are comparable in size with respect to the ground-truth. Earlier in Table \ref{stat}, we have mentioned the number of
communities detected by PVOC based algorithms (the number of communities does not change due to the inclusion of PVOC step with
Louvain and Infomap). In Table \ref{size}, we show for both LFR and real-world networks
that the size of the largest and smallest communities detected by BIGCLAM is much larger than that present in the ground-truth structure. We also measure the similarity (using Jaccard coefficient) between the largest and smallest-size communities detected by BIGCLAM and PVOC based algorithms with the communities in ground-truth structure and notice that PVOC
based algorithms are able to detect both largest and smallest-size communities which are most similar to the ground-truth structure. Therefore,
we hypothesize that our algorithm has the potentiality 
to produce meaningful communities which have high resemblance with the ground-truth structure.

\begin{table}[!h]
\centering
\caption{Size of the largest and smallest communities present in the ground-truth and that obtained from BIGCLAM and PVOC based
algorithms
(the Jaccard similarities between the results obtained from the algorithms with the ground-truth structure are reported within parenthesis) for
both LFR and real-world networks.}\label{size}
\scalebox{0.7}{
 \begin{tabular}{|c|c|c|c|c|c|c|c|c|}
\hline
Networks& \multicolumn{2}{c|}{Ground-truth} & \multicolumn{2}{c|}{BIGCLAM} & \multicolumn{2}{c|}{PVOC+LVN} &
  \multicolumn{2}{c|}{PVOC+INFO}\\\cline{2-9}
               & Max Size & Min size & Max Size & Min size & Max Size & Min size & Max Size & Min size \\\hline   
DBLP           & 3,458 & 124 & 9,876 (0.56) & 877 (0.48) & 4,098 (0.71) & 243 (0.82) & 4,143 (0.76) & 204 (0.81) \\\hline
 Amazon         & 5,987 & 245 & 10,109 (0.45) & 765 (0.57) & 6,876 (0.69) & 398 (0.75) & 6,367 (0.72) & 323 (0.83)\\\hline
 Orkut          & 10,687 & 1,876 & 13,768 (0.72) & 2,985 (0.69) & 11,976 (0.74) & 1,908 (0.79) & 11,345 (0.75) & 1,976 (0.79)\\\hline
Youtube         & 8,987 & 765 & 9,976 (0.65) & 1,098 (0.62) & 8,876 (0.76) & 987 (0.71) & 9,018 (0.74) & 865 (0.82)\\\hline   
 \end{tabular}}

\end{table}

\section{Conclusions}\label{conclusion}
In this paper, we presented a study to show that there is perhaps less need of developing yet another algorithm for finding overlapping
communities from
the network. We demonstrated how the output of an efficient disjoint community detection algorithm can be leveraged  to discover the
overlapping community structure. For that, we proposed a novel, two-phase framework, called PVOC that can be combined with any efficient
disjoint community detection algorithm. PVOC uses a new metric, called permanence in the post-processing step on 
each vertex and detects the  overlapping vertices from the non-overlapping structure. We combined PVOC with two efficient and
scalable algorithms, Louvain and Informap. Experimental results showed that our approach is viable in producing meaningful overlapping
communities quite efficiently even from the large real world networks in terms of high resemblance with the ground-truth community
structure.  PVOC is controlled by only one parameter $\theta$, which can be efficiently tuned to increase the extent of overlapping
memberships per vertex in a network. 

However, a major drawback of PVOC is that it produces exactly the same number of overlapping communities that the disjoint community
detection algorithm produces. However, it might be possible that due to the overlapping nature of a community, new community might emerge
from the disjoint community structure. As an immediate step, we would like to include a new module in the post-processing step that
would consider the emergence of new communities. Moreover, we would try to evaluate PVOC in conjunction with even more disjoint community
detection algorithms.  To conclude, we would like to emphasize on the fact that considering such a massive literature particularly on
community
detection, it is perhaps the good time to put an end to such consistent effort of proposing yet another algorithm, and to revisit
some of the existing algorithms that are efficient enough to fulfill both the purpose of discovering disjoint and overlapping communities
from the network.

\section{Reference}
% \bibliographystyle{iopart-num}
% \bibliography{ref}

\providecommand{\newblock}{}

\if{0}
\section{Introduction: file preparation and submission}

The \verb"iopart" \LaTeXe\ article class file is provided to help authors prepare articles for submission to IOP Publishing journals.
  This document gives advice on preparing your submission, and specific instructions on how to use \verb"iopart.cls" to follow this advice.  You
  do not have to use \verb"iopart.cls"; articles prepared using any other common class and style files can also be submitted.
    It is not necessary to mimic the appearance of a published article.

The advice
on \LaTeX\ file preparation in this document applies to
the journals listed in table~\ref{jlab1}.  If your journal is not listed please go to the journal website via \verb"http://iopscience.iop.org/journals" for specific
submission instructions.

\begin{table}
\caption{\label{jlab1}Journals to which this document applies, and macros for the abbreviated journal names in {\tt iopart.cls}. Macros for other journal titles are listed in appendix\,A.}
\footnotesize
\begin{tabular}{@{}llll}
\br
Short form of journal title&Macro name&Short form of journal title&Macro name\\\hline
2D Mater.&\verb"\TDM"&Mater. Res. Express&\verb"\MRE"\\
Biofabrication&\verb"\BF"&Meas. Sci. Technol.$^c$&\verb"\MST"\\
Bioinspir. Biomim.&\verb"\BB"&Methods Appl. Fluoresc.&\verb"\MAF"\\
Biomed. Mater.&\verb"\BMM"&Modelling Simul. Mater. Sci. Eng.&\verb"\MSMSE"\\
Class. Quantum Grav.&\verb"\CQG"&Nucl. Fusion$^a$&\verb"\NF"\\
Comput. Sci. Disc.&\verb"\CSD"&New J. Phys.&\verb"\NJP"\\
Environ. Res. Lett.&\verb"\ERL"&Nonlinearity$^{a,b}$&\verb"\NL"\\
Eur. J. Phys.$^a$&\verb"\EJP"&Nanotechnology&\verb"\NT"\\
Inverse Problems$^{b,c}$&\verb"\IP"&Phys. Biol.$^c$&\verb"\PB"\\
J. Breath Res.&\verb"\JBR"&Phys. Educ.$^a$&\verb"\PED"\\
J. Geophys. Eng.$^a$&\verb"\JGE"&Physiol. Meas.$^{c,d,e}$&\verb"\PM"\\
J. Micromech. Microeng.&\verb"\JMM"&Phys. Med. Biol.$^{c,d,e}$&\verb"\PMB"\\
J. Neural Eng.$^c$&\verb"\JNE"&Plasma Phys. Control. Fusion&\verb"\PPCF"\\
J. Opt.&\verb"\JOPT"&Phys. Scr.&\verb"\PS"\\
J. Phys. A: Math. Theor.&\verb"\jpa"&Plasma Sources Sci. Technol.&\verb"\PSST"\\
J. Phys. B: At. Mol. Opt. Phys.$^a$&\verb"\jpb"&Rep. Prog. Phys.$^{e}$&\verb"\RPP"\\
J. Phys: Condens. Matter&\verb"\JPCM"&Semicond. Sci. Technol.&\verb"\SST"\\
J. Phys. D: Appl. Phys.$^a$&\verb"\JPD"&Smart Mater. Struct.&\verb"\SMS"\\
J. Phys. G: Nucl. Part. Phys.&\verb"\jpg"&Supercond. Sci. Technol.&\verb"\SUST"\\
J. Radiol. Prot.$^a$&\verb"\JRP"&Surf. Topogr.: Metrol. Prop.&\verb"\STMP"\\
Metrologia$^a$&\verb"\MET"&Transl. Mater. Res.&\verb"\TMR"\\
\br
\end{tabular}\\
$^{a}$UK spelling is required; $^{b}$MSC classification may be used as well as PACS; $^{c}$titles of articles are required in journal references; $^{d}$Harvard-style references must be used (see section \ref{except}); $^{e}$final page numbers of articles are required in journal references.

\end{table}
\normalsize

Any special submission requirements for the journals are indicated with footnotes in table~\ref{jlab1}.
Journals which require references in a particular format will need special care if you are using BibTeX, and you might need to use a \verb".bst" file
that gives slightly non-standard output in order to supply any extra information required.  It is not
necessary to give references in the exact style of references used in published articles, as long as all of
the required information is present.

Also note that there is an incompatibility
between \verb"amsmath.sty" and \verb"iopart.cls" which cannot be completely worked around.  If your article relies
on commands in \verb"amsmath.sty" that are not available in \verb"iopart.cls", you may wish to consider using a different
class file.

Whatever journal you are submitting to, please look at recent published articles (preferably
articles in your subject area) to familiarize yourself with the features of the journal.  We do not demand
that your \LaTeX\ file closely resembles a published article---a generic `preprint' appearance of the sort
commonly seen on \verb"arXiv.org" is fine---but your submission should be presented
in a way that makes it easy for the referees to form an opinion of whether it is suitable for the journal.
The generic advice in this document---on what to include in an abstract, how best to present complicated
mathematical expressions, and so on---applies whatever class file you are using.

\subsection{What you will need to supply}
Submissions to our journals are handled via the ScholarOne web-based submission system.  When you submit
a new article to us you need only submit a PDF of your article.  When you submit a revised version,
we ask you to submit the source files as well.  Upon acceptance for publication we will use the source files to produce a proof of your article in the journal style. 

\subsubsection{Text.}When you send us the source files for a revised version of your submission,
you should send us the \LaTeX\ source code of your paper with all figures read in by 
the source code (see section \ref{figinc}).  Articles can be prepared using almost any version of \TeX\ or \LaTeX{},
not just \LaTeX\ with the class file \verb"iopart.cls".  You may split your \LaTeX\ file into several parts, but please show
which is the `master' \LaTeX\ file that reads in all of the other ones by naming it appropriately.  The `master'
\LaTeX\ file must read in all other \LaTeX\ and figure files from the current directory.  {\it Do not read in files from a different directory, e.g. \verb"\includegraphics{/figures/figure1.eps}" or
\verb"\include{../usr/home/smith/myfiles/macros.tex}"---we store submitted files
all together in a single directory with no subdirectories}.
\begin{itemize}
\item {\bf Using \LaTeX\ packages.} Most \LaTeXe\ packages can be used if they are 
available in common distributions of \LaTeXe; however, if it is essential to use 
a non-standard package then any extra files needed to process the article must 
also be supplied.  Try to avoid using any packages that manipulate or change the standard
\LaTeX\ fonts: published articles use fonts in the Times family, but we prefer that you 
use \LaTeX\ default Computer Modern fonts in your submission.  The use of \LaTeX\ 2.09, and of plain
\TeX\ and variants such as AMSTeX is acceptable, but a complete PDF of your submission should be supplied in these cases.
\end{itemize}
\subsubsection{Figures.} Figures should ideally be included in an article as encapsulated PostScript files
(see section \ref{figinc}) or created using standard \LaTeX\ drawing commands. 
 Please name all figure files using the guidelines in section \ref{fname}.
We accept submissions that use pdf\TeX\ to include
PDF or bitmap figures, but please ensure that you send us a PDF that uses PDF version 1.4 or lower
(to avoid problems in the ScholarOne system).
You can do this by putting \verb"\pdfminorversion=4" at the very start of your TeX file.

\label{fig1}All figures should be included within the body of the text 
at an appropriate point or grouped together with their captions at the end of the article. A standard graphics inclusion package such as \verb"graphicx" should be used for figure inclusion, and the package should be declared in the usual
way, for example with \verb"\usepackage{graphicx}", after the \verb"\documentclass" command.
Authors should avoid using special effects generated by including verbatim
PostScript code in the submitted \LaTeX\ file. Wherever possible, please try to use standard \LaTeX\ tools 
and packages.

\subsubsection{References.\label{bibby}}
You can produce your bibliography in the standard \LaTeX\ way using the \verb"\bibitem" command. Alternatively
you can use BibTeX: our preferred  \verb".bst" styles are: 

\begin{itemize}
\item For the numerical (Vancouver) reference style we recommend that authors use 
 \verb"unsrt.bst"; this does not quite follow the style of published articles in our
 journals but this is not a problem.  Alternatively \verb"iopart-num.bst" created by Mark A Caprio
 produces a reference style that closely matches that in published articles.  The file is available from
\verb"http://ctan.org/tex-archive/biblio/bibtex/contrib/iopart-num/" .
\item For alphabetical (Harvard) style references we recommend that authors use the \verb"harvard.sty"
in conjunction with the \verb"jphysicsB.bst" BibTeX style file.  These, and accompanying documentation, can be downloaded
from \penalty-10000 \verb"http://www.ctan.org/tex-archive/macros/latex/contrib/harvard/".
Note that the \verb"jphysicsB.bst" bibliography style does not include article titles
in references to journal articles.
To include the titles of journal articles you can use the style \verb"dcu.bst" which is included
in the \verb"harvard.sty" package.  The output differs a little from the final journal reference
style, but all of the necessary information is present and the reference list will be formatted
into journal house style as part of the production process if your article is accepted for publication.
\end{itemize}

\noindent Please make sure that you include your \verb".bib" bibliographic database file(s) and any 
\verb".bst" style file(s) you have used.

\subsection{\label{copyright}Copyrighted material and ethical policy} If you wish to make use of previously published material for which you do not own the copyright then you must seek permission from the copyright holder, usually both the author and the publisher.  It is your responsibility to obtain copyright permissions and this should be done prior to submitting your article. If you have obtained permission, please provide full details of the permission granted---for example, copies of the text of any e-mails or a copy of any letters you may have received. Figure captions must include an acknowledgment of the original source of the material even when permission to reuse has been obtained.  Please read our ethical policy (available at \verb"http://authors.iop.org/ethicalpolicy") before writing your article.

\subsection{Naming your files}
\subsubsection{General.}
Please name all your files, both figures and text, as follows:
\begin{itemize}
\item Use only characters from the set a to z, A to Z, 0 to 9 and underscore (\_).
\item Do not use spaces or punctuation characters in file names.
\item Do not use any accented characters such as
\'a, \^e, \~n, \"o.
\item Include an extension to indicate the file type (e.g., \verb".tex", \verb".eps", \verb".txt", etc).
\item Use consistent upper and lower case in filenames and in your \LaTeX\ file.
If your \LaTeX\ file contains the line \verb"\includegraphics{fig1.eps}" the figure file must be called
\verb"fig1.eps" and not \verb"Fig1.eps" or \verb"fig1.EPS".  If you are on a Unix system, please ensure that
there are no pairs of figures whose names differ only in capitalization, such as \verb"fig_2a.eps" and \verb"fig_2A.eps",
as Windows systems will be unable to keep the two files in the same directory.
\end{itemize}
When you submit your article files, they are manipulated
and copied many times across multiple databases and file systems. Including non-standard
characters in your filenames will cause problems when processing your article.
\subsubsection{\label{fname}Naming your figure files.} In addition to the above points, please give each figure file a name which indicates the number of the figure it contains; for example, \verb"figure1.eps", \verb"figure2a.eps", etc. If the figure file contains a figure with multiple parts, for example figure 2(a) to 2(e), give it a name such as \verb"figure2a_2e.eps", and so forth.
\subsection{How to send your files}
Please send your submission via the ScholarOne submission system.  Go to the journal home
page, and use the 'Submit an article' link on the right-hand side.

\subsection{Obtaining the class file and documentation}
This documentation and the \verb"iopart.cls" class file and associated files are available by anonymous FTP
at \verb"ftp://ftp.iop.org/pub/journals/".  They are available in two archive
formats there, \verb"ioplatexguidelines.tar.gz" and \verb"ioplatexguidelines.zip".  The individual files
are present in the subdirectory \verb"latex2e".

\section{Preparing your article}

\subsection{Sample coding for the start of an article}
\label{startsample}
The code for the start of a title page of a typical paper in the \verb"iopart.cls" style might read:
\small\begin{verbatim}
\documentclass[12pt]{iopart}
\begin{document}
\title[The anomalous magnetic moment of the 
neutrino]{The anomalous magnetic moment of the 
neutrino and its relation to the solar neutrino problem}

\author{P J Smith$^1$, T M Collins$^2$, 
R J Jones$^3$\footnote{Present address:
Department of Physics, University of Bristol, Tyndalls Park Road, 
Bristol BS8 1TS, UK.} and Janet Williams$^3$}

\address{$^1$ Mathematics Faculty, Open University, 
Milton Keynes MK7~6AA, UK}
\address{$^2$ Department of Mathematics, 
Imperial College, Prince Consort Road, London SW7~2BZ, UK}
\address{$^3$ Department of Computer Science, 
University College London, Gower Street, London WC1E~6BT, UK}
\ead{williams@ucl.ac.uk}

\begin{abstract}
...
\end{abstract}
\pacs{1315, 9440T}
\keywords{magnetic moment, solar neutrinos, astrophysics}
\submitto{\jpg}
\maketitle
\end{verbatim}
\normalsize

At the start of the \LaTeX\ source code please include 
commented material to identify the journal, author, and (if you are sending a revised
version or a resubmission) the reference number that the journal
has given to the submission. The first non-commented line should be 
\verb"\documentclass[12pt]{iopart}"  to load the preprint class 
file.  The normal text will be in the Computer Modern 12pt font.
It is possible to specify 10pt font size by passing the option \verb"[10pt]" to the class file.
Although it is possible to choose a font other than Computer Modern by loading external packages, this is not recommended.

The article text begins after \verb"\begin{document}".
Authors of very long articles may find it convenient to separate 
their article into a series of \LaTeX\ files each containing one section, and each of which is called 
in turn by the primary file.  The files for each section should be read in from the current directory;
please name the primary file clearly so that we know to run \LaTeX\ on this file.

Authors may use any common \LaTeX\ \verb".sty" files.
Authors may also define their own macros and definitions either in the main article \LaTeX\ file
or in a separate \verb".tex" or \verb".sty" file that is read in by the
main file, provided they do not overwrite existing definitions.
It is helpful to the production staff if complicated author-defined macros are explained in a \LaTeX\ comment.
The article class \verb"iopart.cls" can be used with other package files such
as those loading the AMS extension fonts 
\verb"msam" and \verb"msbm", which provide the 
blackboard bold alphabet and various extra maths symbols as well as symbols useful in figure 
captions.  An extra style file \verb"iopams.sty" is provided to load these
packages and provide extra definitions for bold Greek letters.

\subsection{\label{dblcol}Double-column layout}
The \verb"iopart.cls" class file produces single-column output by default, but a two-column layout can be obtained by
using \verb"\documentclass[10pt]" at the start of the file and \verb"\ioptwocol" after the \verb"\maketitle" command.  Two-column output will begin
on a new page (unlike in published double-column articles, where the two-column material
starts on the same page as the abstract).

In general we prefer to receive submissions in single-column format even for journals
published in double-column style; however, the \verb"\ioptwocol" option may be useful to test figure sizes
and equation breaks for these journals.  When setting material
in two columns you can use the asterisked versions of \LaTeX\ commands such as \verb"\begin{figure*} ... \end{figure*}"
to set figures and tables across two columns.  If you have any problems or any queries about producing two-column output, please contact us at \verb"submissions@iop.org".

\section{The title and abstract page} 
If you use \verb"iopart.cls", the code for setting the title page information is slightly different from
the normal default in \LaTeX.  If you are using a different class file, you do not need to mimic the appearance of
an \verb"iopart.cls" title page, but please ensure that all of the necessary information is present.

\subsection{Titles and article types}
The title is set using the command
\verb"\title{#1}", where \verb"#1" is the title of the article. The
first letter 
of the title should be capitalized with the rest in lower case. 
The title appears in bold case, but mathematical expressions within the title may be left in light-face type. 

If the title is too long to use as a running head at the top of each page (apart from the
first) a short
form can be provided as an optional argument (in square brackets)
before the full title, i.e.\ \verb"\title[Short title]{Full title}".

For article types other than papers, \verb"iopart.cls"
has a generic heading \verb"\article[Short title]{TYPE}{Full title}" 
and some specific definitions given in table~\ref{arttype}. In each case (apart from Letters
to the Editor and Fast Track Communications) an 
optional argument can be used immediately after the control sequence name
to specify the short title; where no short title is given, the full title
will be used as the running head.  Not every article type has its own macro---use \verb"\article" for
any not listed.  A full list of the types of articles published by a journal is given
in the submission information available via the journal home page.
%For Letters use \verb"\letter{Full title}"; no short title is required as 
%the running head is automatically defined to be {\it Letter to the Editor}.
The generic heading could be used for 
articles such as those presented at a conference or workshop, e.g.
\small\begin{verbatim}
\article[Short title]{Workshop on High-Energy Physics}{Title}
\end{verbatim}\normalsize
Footnotes to titles may be given by using \verb"\footnote{Text of footnote.}" immediately after the title.
Acknowledgment of funding should be included in the acknowledgments section rather than in a footnote.

\begin{table}
\caption{\label{arttype}Types of article defined in the {\tt iopart.cls} 
class file.}
\footnotesize\rm
\begin{tabular*}{\textwidth}{@{}l*{15}{@{\extracolsep{0pt plus12pt}}l}}
\br
Command& Article type\\
\mr
\verb"\title{#1}"&Paper (no surtitle on first page)\\
\verb"\ftc{#1}"&Fast Track Communication\\
\verb"\review{#1}"&Review\\
\verb"\topical{#1}"&Topical Review\\
\verb"\comment{#1}"&Comment\\
\verb"\note{#1}"&Note\\
\verb"\paper{#1}"&Paper (no surtitle on first page)\\
\verb"\prelim{#1}"&Preliminary Communication\\
\verb"\rapid{#1}"&Rapid Communication\\
\verb"\letter{#1}"&Letter to the Editor\\
\verb"\article{#1}{#2}"&Other articles\\\ & (use this for any other type of article; surtitle is whatever is entered as {\tt 
\#1})\\
\br
\end{tabular*}
\end{table}

\subsection{Authors' names and addresses}
For the authors' names type \verb"\author{#1}", 
where \verb"#1" is the 
list of all authors' names. Western-style names should be written as initials then
family name, with a comma after all but the last 
two names, which are separated by `and'. Initials should {\it not} be followed by full stops. First (given) names may be used if 
desired.  Names in Chinese, Japanese and Korean styles should be written as you want them to appear in the published article. Authors in all IOP Publishing journals have the option to include their names in Chinese, Japanese or Korean characters in addition to the English name: see appendix B for details.

If the authors are at different addresses a superscripted number, e.g. $^1$, \verb"$^1$", should be used after each 
name to reference the author to his/her address.
If an author has additional information to appear as a footnote, such as 
a permanent address, a normal \LaTeX\ footnote command
should be given after the family name and address marker 
with this extra information.

The authors' affiliations follow the list of authors. 
Each address is set by using
\verb"\address{#1}" with the address as the single parameter in braces. 
If there is more 
than one address then the appropriate superscripted number, followed by a space, should come at the start of
the address.
 
E-mail addresses are added by inserting the 
command \verb"\ead{#1}" after the postal address(es) where \verb"#1" is the e-mail address.  
See section~\ref{startsample} for sample coding. For more than one e-mail address, please use the command 
\verb"\eads{\mailto{#1}, \mailto{#2}}" with \verb"\mailto" surrounding each e-mail address.  Please ensure
that, at the very least, you state the e-mail address of the corresponding author.

\subsection{The abstract}
The abstract follows the addresses and
should give readers concise information about the content 
of the article and indicate the main results obtained and conclusions 
drawn. It should be self-contained---there should be no references to 
figures, tables, equations, bibliographic references etc.  It should be enclosed between \verb"\begin{abstract}"
and \verb"\end{abstract}" commands.  The abstract should normally be restricted 
to a single paragraph of around 200 words.

\subsection{Subject classification numbers}
For all of our journals, we ask that you should state Physics and Astronomy Classification System (PACS)
classification numbers, except for {\it Inverse Problems} and {\it Nonlinearity} whose authors may 
use either PACS or Mathematics Subject Classification (MSC) codes.  These codes
can greatly help in the choice of suitable referees and allocation of articles to subject areas.
See \verb"http://www.aip.org/pacs" and \verb"http://www.ams.org/msc" 
for full information on these codes.  

PACS or MSC numbers are included after the abstract 
using \verb"\pacs{#1}" and \verb"\ams{#1}" respectively.

After any classification numbers the command
\verb"\submitto{#1}" can be inserted, where \verb"#1" is the journal name written in full or the appropriate control sequence as
given in table~\ref{jlab1}. This command is not essential to the running of the file and can be omitted.

\subsection{Keywords}
Keywords are required for all submissions. Authors should supply a minimum of three (maximum seven) keywords appropriate to their article as a new paragraph starting \verb"\noindent{\it Keywords\/}:" after the end of the abstract.

\subsection{Making a separate title page}
To keep the header material on a separate page from the
body of the text insert \verb"\maketitle" (or \verb"\newpage") before the start of the text. 
If \verb"\maketitle" is not included the text of the
article will start immediately after the abstract.  

\section{The text}
\subsection{Sections, subsections and subsubsections}
The text of articles may be divided into sections, subsections and, where necessary, 
subsubsections. To start a new section, end the previous paragraph and 
then include \verb"\section" followed by the section heading within braces. 
Numbering of sections is done {\it automatically} in the headings: 
sections will be numbered 1, 2, 3, etc, subsections will be numbered 
2.1, 2.2,  3.1, etc, and subsubsections will be numbered 2.3.1, 2.3.2, 
etc.  Cross references to other sections in the text should, where
possible, be made using 
labels (see section~\ref{xrefs}) but can also
be made manually. See section~\ref{eqnum} for information on the numbering of displayed equations. Subsections and subsubsections are 
similar to sections but 
the commands are \verb"\subsection" and \verb"\subsubsection" respectively. 
Sections have a bold heading, subsections an italic heading and 
subsubsections an italic heading with the text following on directly.
\small\begin{verbatim}
\section{This is the section title}
\subsection{This is the subsection title}
\end{verbatim}\normalsize

The first section is normally an introduction,  which should state clearly 
the object of the work, its scope and the main advances reported, with 
brief references to relevant results by other workers. In long papers it is 
helpful to indicate the way in which the paper is arranged and the results 
presented.

Footnotes should be avoided whenever possible and can often be included in the text as phrases or sentences in parentheses. If required, they should be used only for brief notes that do not fit conveniently into the text. The use of 
displayed mathematics in footnotes should be avoided wherever possible and no equations within a footnote should be numbered. 
The standard \LaTeX\ macro \verb"\footnote" should be used.  Note that in \verb"iopart.cls" the \verb"\footnote" command
produces footnotes indexed by a variety of different symbols,
whereas in published articles we use numbered footnotes.  This
is not a problem: we will convert symbol-indexed footnotes to numbered ones during the production process.

\subsection{Acknowledgments}
Authors wishing to acknowledge assistance or encouragement from 
colleagues, special work by technical staff or financial support from 
organizations should do so in an unnumbered `Acknowledgments' section 
immediately following the last numbered section of the paper. In \verb"iopart.cls" the 
command \verb"\ack" sets the acknowledgments heading as an unnumbered
section.

Please ensure that you include all of the sources of funding and the funding contract reference numbers that you are contractually obliged to acknowledge. We often receive requests to add such information very late in the production process, or even after the article is published, and we cannot always do this. Please collect all of the necessary information from your co-authors and sponsors as early as possible.  

\subsection{Appendices}
Technical detail that it is necessary to include, but that interrupts 
the flow of the article, may be consigned to an appendix. 
Any appendices should be included at the end of the main text of the paper, after the acknowledgments section (if any) but before the reference list.
If there are 
two or more appendices they should be called Appendix A, Appendix B, etc. 
Numbered equations will be in the form (A.1), (A.2), etc,
figures will appear as figure A1, figure B1, etc and tables as table A1,
table B1, etc.

The command \verb"\appendix" is used to signify the start of the
appendices. Thereafter \verb"\section", \verb"\subsection", etc, will 
give headings appropriate for an appendix. To obtain a simple heading of 
`Appendix' use the code \verb"\section*{Appendix}". If it contains
numbered equations, figures or tables the command \verb"\appendix" should
precede it and \verb"\setcounter{section}{1}" must follow it. 
 
\subsection{Some matters of style}
It will help the readers if your article is written in a clear,
consistent and concise manner. During the production process
we will try to make sure that your work is presented to its
readers in the best possible way without sacrificing the individuality of
your writing. Some recommended 
points to note, however, are the following.  These apply to all of the journals listed
in table~\ref{jlab1}.
\begin{enumerate}
\item Authors are often inconsistent in the use of `ize' and `ise' endings.
We recommend using `-ize' spellings (diagonalize, 
renormalization, minimization, etc) but there are some common 
exceptions to this, for example: devise, 
promise and advise.

\item The words table and figure should be written 
in full and {\bf not} abbreviaged to tab. and fig. Do not include `eq.', `equation' etc before an equation number or `ref.'\, `reference' etc before a reference number.
\end{enumerate}

Please check your article carefully for accuracy, consistency and clarity before
submission. Remember that your article will probably be read by many
people whose native language is not English and who may not  
be aware of many of the subtle meanings of words or idiomatic phases
present in the English language. It therefore helps if you try to keep
sentences as short and simple as possible.  If you are not a native English speaker,
please ask a native English speaker to read your paper and check its grammar.

\section{Mathematics}
\subsection{Two-line constructions}
The great advantage of \LaTeX\ 
over other text processing systems is its 
ability to handle mathematics of almost any degree of complexity. However, 
in order to produce an article suitable for publication both within a print journal and online, 
authors should exercise some restraint on the constructions used. Some equations using very small characters which are clear in a preprint style article may be difficult read in a smaller format.

For simple fractions in the text the solidus \verb"/", as in 
$\lambda/2\pi$, should be used instead of \verb"\frac" or \verb"\over", 
using parentheses where necessary to avoid ambiguity, 
for example to distinguish between $1/(n-1)$ and $1/n-1$. Exceptions to 
this are the proper fractions $\frac12$, $\frac13$, $\frac34$, 
etc, which are better left in this form. In displayed equations 
horizontal lines are preferable to solidi provided the equation is 
kept within a height of two lines. A two-line solidus should be 
avoided where possible; the construction $(\ldots)^{-1}$ should be 
used instead. For example use:
\begin{equation*}
\frac{1}{M_{\rm a}}\left(\int^\infty_0{\rm d}
\omega\;\frac{|S_o|^2}{N}\right)^{-1}\qquad\mbox{instead of}\qquad
\frac{1}{M_{\rm a}}\biggl/\int^\infty_0{\rm d}
\omega\;\frac{|S_o|^2}{N}.
\end{equation*}

\subsection{Roman and italic in mathematics}
In mathematics mode \LaTeX\ automatically sets variables in an italic 
font. In most cases authors should accept this italicization. However, 
there are some cases where it is preferable to use a Roman font; for 
instance, a Roman d for a differential d, a Roman e 
for an exponential e and a Roman i for the square root of $-1$. To 
accommodate this and to simplify the  typing of equations, \verb"iopart.cls" provides
some extra definitions. \verb"\rmd", \verb"\rme" and \verb"\rmi" 
now give Roman d, e and i respectively for use in equations, 
e.g.\ $\rmi x\rme^{2x}\rmd x/\rmd y$ 
is obtained by typing \verb"$\rmi x\rme^{2x}\rmd x/\rmd y$".

Certain other common mathematical functions, such as cos, sin, det and 
ker, should appear in Roman type. Standard \LaTeX\ provides macros for 
most of these functions 
(in the cases above, \verb"\cos", \verb"\sin", \verb"\det" and \verb"\ker" 
respectively); \verb"iopart.cls" also provides 
additional definitions for $\Tr$, $\tr$ and 
$\Or$ (\verb"\Tr", \verb"\tr" and \verb"\Or", respectively). 

Subscripts and superscripts should be in Roman type if they are labels 
rather than variables or characters that take values. For example in the 
equation
\[
\epsilon_m=-g\mu_{\rm n}Bm
\]
$m$, the $z$ component of the nuclear spin, is italic because it can have 
different values whereas n is Roman because it 
is a label meaning nuclear ($\mu_{\rm n}$ 
is the nuclear magneton).

\subsection{Displayed equations in double-column journals}
Authors should bear in mind that all mathematical formulae in double-column journals will need to fit
into the width of a single column.  You may find it helpful to use a two-column layout (such as the two-column
option in \verb"iopart.cls") in your submission so that you can check the width of equations.

\subsection{Special characters for mathematics}
Bold italic characters can be used in our journals to signify vectors (rather
than using an upright bold or an over arrow). To obtain this effect when using \verb"iopart.cls",
use \verb"\bi{#1}" within maths mode, e.g. $\bi{ABCdef}$. Similarly, in \verb"iopart.cls", if upright 
bold characters are required in maths, use \verb"\mathbf{#1}" within maths
mode, e.g. $\mathbf{XYZabc}$. The calligraphic (script) uppercase alphabet
is obtained with \verb"\mathcal{AB}" or \verb"\cal{CD}" 
($\mathcal{AB}\cal{CD}$).

The American Mathematical Society provides a series of extra symbol fonts
to use with \LaTeX\ and packages containing the character definitions to
use these fonts. Authors wishing to use Fraktur 
\ifiopams$\mathfrak{ABC}$ \fi
or Blackboard Bold \ifiopams$\mathbb{XYZ}$ \fi can include the appropriate
AMS package (e.g. \verb"amsgen.sty", \verb"amsfonts.sty", \verb"amsbsy.sty", \verb"amssymb.sty") with a 
\verb"\usepackage" command or add the command \verb"\usepackage{iopams}"
which loads the four AMS packages mentioned above and also provides
definitions for extra bold characters (all Greek letters and some other
additional symbols). 

The package \verb"iopams.sty" uses the definition \verb"\boldsymbol" in \verb"amsbsy.sty"
which allows individual non-alphabetical symbols and Greek letters to be 
made bold within equations.
The bold Greek lowercase letters \ifiopams$\balpha \ldots\bomega$,\fi 
are obtained with the commands 
\verb"\balpha" \dots\ \verb"\bomega" (but note that
bold eta\ifiopams, $\bfeta$,\fi\ is \verb"\bfeta" rather than \verb"\beta")
and the capitals\ifiopams, $\bGamma\ldots\bOmega$,\fi\ with commands 
\verb"\bGamma" \dots\
\verb"\bOmega". Bold versions of the following symbols are
predefined in \verb"iopams.sty": 
bold partial\ifiopams, $\bpartial$,\fi\ \verb"\bpartial",
bold `ell'\ifiopams, $\bell$,\fi\  \verb"\bell", 
bold imath\ifiopams, $\bimath$,\fi\  \verb"\bimath", 
bold jmath\ifiopams, $\bjmath$,\fi\  \verb"\bjmath", 
bold infinity\ifiopams, $\binfty$,\fi\ \verb"\binfty", 
bold nabla\ifiopams, $\bnabla$,\fi\ \verb"\bnabla", 
bold centred dot\ifiopams, $\bdot$,\fi\  \verb"\bdot". Other 
characters are made bold using 
\verb"\boldsymbol{\symbolname}".

Please do not use the style file \verb"amsmath.sty" (part of the AMSTeX package) in conjunction with \verb"iopart.cls". This will result in several errors. To make use of the macros defined in \verb"amsmath.sty", \verb"iopart.cls" provides the file \verb"setstack.sty" which reproduces the following useful macros from \verb"amsmath.sty":
\small\begin{verbatim}
\overset \underset \sideset \substack \boxed   \leftroot
\uproot  \dddot    \ddddot  \varrow   \harrow
\end{verbatim}\normalsize

If the mathematical notation
that you need is best handled in \verb"amsmath.sty" you might want to consider using an article class
other than \verb"iopart.cls". We accept submissions using any class or style files.

Table~\ref{math-tab2} lists some other macros for use in 
mathematics with a brief description of their purpose.

\begin{table}
\caption{\label{math-tab2}Other macros defined in {\tt iopart.cls} for use in maths.}
\begin{tabular*}{\textwidth}{@{}l*{15}{@{\extracolsep{0pt plus
12pt}}l}}
\br
Macro&Result&Description\\
\mr
\verb"\fl"&&Start line of equation full left\\
\verb"\case{#1}{#2}"&$\case{\#1}{\#2}$&Text style fraction in display\\
\verb"\Tr"&$\Tr$&Roman Tr (Trace)\\
\verb"\tr"&$\tr$&Roman tr (trace)\\
\verb"\Or"&$\Or$&Roman O (of order of)\\
\verb"\tdot{#1}"&$\tdot{x}$&Triple dot over character\\
\verb"\lshad"&$\lshad$&Text size left shadow bracket\\
\verb"\rshad"&$\rshad$&Text size right shadow bracket\\
\br
\end{tabular*}
\end{table}

\subsection{Alignment of displayed equations}

The normal style for aligning displayed equations in our published journal articles is to align them left rather than centre. The \verb"iopart.cls" class file automatically does this and indents each line of a display.  In \verb"iopart.cls", to make any line start at the left margin of the page, add \verb"\fl" at start of the line (to indicate full left).

Using the \verb"eqnarray" environment equations will naturally be aligned left and indented without the use of any ampersands for alignment, see equations (\ref{eq1}) and (\ref{eq2})
\begin{eqnarray}
\alpha + \beta =\gamma^2, \label{eq1}\\
\alpha^2 + 2\gamma + \cos\theta = \delta. \label{eq2} 
\end{eqnarray}
This is the normal equation style for our journals.

Where some secondary alignment is needed, for instance a second part of an equation on a second line, a single ampersand is added at the point of alignment in each line  (see  (\ref{eq3}) and (\ref{eq4})).
\begin{eqnarray}
\alpha &=2\gamma^2 + \cos\theta + \frac{XY \sin\theta}{X+ Y\cos\theta} \label{eq3}\\
 & = \delta\theta PQ \cos\gamma. \label{eq4} 
\end{eqnarray}
 
Two points of alignment are possible using two ampersands for alignment (see  (\ref{eq5}) and (\ref{eq6})).  Note in this case extra space \verb"\qquad" is added before the second ampersand in the longest line (the top one) to separate the condition from the equation. 
\begin{eqnarray}
\alpha &=2\gamma^2 + \cos\theta + \frac{XY \sin\theta}{X+ Y\cos\theta}\qquad& \theta > 1 \label{eq5}\\
 & = \delta\theta PQ \cos\gamma &\theta \leq 1.\label{eq6} 
\end{eqnarray}

For a long equation which has to be split over more than one line the first line should start at the left margin, this is achieved by inserting \verb"\fl" (full left) at the start of the line. The use of the alignment parameter \verb"&" is not necessary unless some secondary alignment is needed.
\begin{eqnarray}
\fl \alpha + 2\gamma^2 = \cos\theta + \frac{XY \sin\theta}{X+ Y\cos\theta} +  \frac{XY \sin\theta}{X- Y\cos\theta} +
+ \left(\frac{XY \sin\theta}{X+ Y\cos\theta}\right)^2 \nonumber\\
+  \left(\frac{XY \sin\theta}{X- Y\cos\theta}\right)^2.\label{eq7} 
\end{eqnarray}

The plain \TeX\ command \verb"\eqalign" can be used within an \verb"equation" environment to obtain a multiline equation with a single centred number, for example
\begin{equation}
\eqalign{\alpha + \beta =\gamma^2 \cr
\alpha^2 + 2\gamma + \cos\theta = \delta.} 
\end{equation}

During the production process we will break equations as appropriate for the page layout of the journal. If you are submitting to a double-column journal and wish to review how your equations will break, you may find the double-column layout described in section \ref{dblcol} useful.
 
\subsection{Miscellaneous points}
The following points on the layout of mathematics apply whichever class file you use.

Exponential expressions, especially those containing subscripts or 
superscripts, are clearer if the notation $\exp(\ldots)$ is used, except for 
simple examples. For instance $\exp[\rmi(kx-\omega t)]$ and $\exp(z^2)$ are 
preferred to $\e^{\rmi(kx-\omega t)}$ and $\e^{z^2}$, but 
$\e^x$ 
is acceptable. 

Similarly the square root sign $\sqrt{\phantom{b}}$ should 
only be used with relatively
simple expressions, e.g.\ $\sqrt2$ and $\sqrt{a^2+b^2}$;
in other cases the 
power $1/2$ should be used; for example, $[(x^2+y^2)/xy(x-y)]^{1/2}$.

It is important to distinguish between $\ln = \log_\e$ and $\lg 
=\log_{10}$. Braces, brackets and parentheses should be used in the 
following order: $\{[(\;)]\}$. The same ordering of brackets should be 
used within each size. However, this ordering can be ignored if the
brackets have a 
special meaning (e.g.\ if they denote an average or a function).  

Decimal fractions should always be preceded by a zero: for example 0.123 {\bf not} .123.
For long numbers use thin spaces after every third character away from the position of the decimal point, unless 
this leaves a single separated character: e.g.\ $60\,000$, $0.123\,456\,78$ 
but 4321 and 0.7325.

Equations should be followed by a full stop (periods) when at the end
of a sentence.

\subsection{Equation numbering and layout in {\tt iopart.cls}}
\label{eqnum}
\LaTeX\ provides facilities for automatically numbering equations 
and these should be used where possible. Sequential numbering (1), (2), 
etc, is the default numbering system although in \verb"iopart.cls", if the command
\verb"\eqnobysec" is included in the preamble, equation numbering
by section is obtained, e.g.\ 
(2.1), (2.2), etc. Equation numbering by section is used in appendices automatically when the \verb"\appendix" command is used, even if sequential numbering has been used in the rest of the article. 
Refer to equations in the text using the equation number in parentheses. It is not normally necessary to include the word equation before the number; and abbreviations such as eqn or eq should not be used.
In \verb"iopart.cls", there are alternatives to the standard \verb"\ref" command that you might
find useful---see \tref{abrefs}.

Sometimes it is useful to number equations as parts of the same
basic equation. This can be accomplished in \verb"iopart.cls" by inserting the 
commands \verb"\numparts" before the equations concerned and 
\verb"\endnumparts" when reverting to the normal sequential numbering.
For example using \verb"\numparts \begin{eqnarray}" ... \verb"\end{eqnarray} \endnumparts":

\numparts
\begin{eqnarray}
T_{11}&=(1+P_\e)I_{\uparrow\uparrow}-(1-P_\e)
I_{\uparrow\downarrow},\label{second}\\
T_{-1-1}&=(1+P_\e)I_{\downarrow\downarrow}-(1-P_\e)I_{\uparrow\downarrow},\\
S_{11}&=(3+P_\e)I_{\downarrow\uparrow}-(3-P_e)I_{\uparrow\uparrow},\\
S_{-1-1}&=(3+P_\e)I_{\uparrow\downarrow}-(3-P_\e)
I_{\downarrow\downarrow}.
\end{eqnarray}
\endnumparts

Equation labels within the \verb"\eqnarray" environment will be referenced
as subequations, e.g. (\ref{second}).

\subsection{Miscellaneous extra commands for displayed equations}
The \verb"\cases" command has been amended slightly in \verb"iopart.cls" to 
increase the space between the equation and the condition. 
\Eref{cases} 
demonstrates simply the output from the \verb"\cases" command
\begin{equation}
\label{cases}
X=\cases{1&for $x \ge 0$\\
-1&for $x<0$\\}
\end{equation}
%The code used was:
%\small\begin{verbatim}
%\begin{equation}
%\label{cases}
%X=\cases{1&for $x \ge 0$\\
%-1&for $x<0$\\}
%\end{equation}
%\end{verbatim}
%\normalsize

To obtain text style fractions within displayed maths the command 
\verb"\case{#1}{#2}" can be used instead
of the usual \verb"\frac{#1}{#2}" command or \verb"{#1 \over #2}".

When two or more short equations are on the same line they should be 
separated by a `qquad space' (\verb"\qquad"), rather than
\verb"\quad" or any combination of \verb"\,", \verb"\>", \verb"\;" 
and \verb"\ ".

\section{Referencing\label{except}}
Two different styles of referencing are in common use: 
the Harvard alphabetical system and the Vancouver numerical system. 
All journals to which this document applies allow the use of either the Harvard or Vancouver system, 
except for {\it Physics in Medicine and Biology} and {\it Physiological Measurement}
for which authors {\it must\/} use the Harvard referencing style (with the titles of journal
articles given, and final page numbers given). 

\subsection{Harvard (alphabetical) system}
In the Harvard system the name of the author appears in the text together 
with the year of publication. As appropriate, either the date or the name 
and date are included within parentheses. Where there are only two authors 
both names should be given in the text; if there are more than two 
authors only the first name should appear followed by `{\it et al}' 
(which can be obtained in \verb"iopart.cls" by typing \verb"\etal"). When two or 
more references to work by one author or group of authors occur for the 
same year they should be identified by including a, b, etc after the date 
(e.g.\ 2012a). If several references to different pages of the same article 
occur the appropriate page number may be given in the text, e.g.\ Kitchen 
(2011, p 39).

The reference list at the end of an article consists of an 
unnumbered `References' section containing an
alphabetical listing by authors' names. References with the same author list are ordered by date, oldest first.
The reference list in the 
preprint style is started in \verb"iopart.cls" by including the command \verb"\section*{References}" and then
\verb"\begin{harvard}".
Individual references start with \verb"\item[]" and the reference list is completed with \verb"\end{harvard}".
There is also a shortened form of the coding: \verb"\section*{References}"
and \verb"\begin{harvard}" can be replaced by the single command
\verb"\References", and \verb"\end{harvard}" can be shortened to
\verb"\endrefs".

\subsection{Vancouver (numerical) system}
In the Vancouver system references are numbered sequentially 
throughout the text. The numbers occur within square brackets and one 
number can be used to designate several references. A numerical 
reference list in the \verb"iopart" style is started by including the 
command \verb"\section*{References}" and then
\verb"\begin{thebibliography}{<num>}", where \verb"<num>" is the largest
number in the reference list (or any other number with the same number
of digits).  The 
reference list gives the references in 
numerical order, individual references start with \verb"\bibitem{label}". The list is completed by
\verb"\end{thebibliography}". Short forms of the commands are again
available: \verb"\Bibliography{<num>}" can be used at the start of the
references section and \verb"\endbib" at the end.

A variant of this system is to use labels instead of numbers within 
square brackets, in this case references in the list should start with \verb"\bibitem[label-text]". This method is allowed for all journals that accept numerical references.

\subsection{BibTeX\label{bibtex}}
If you are using BibTeX, see the earlier section \ref{bibby} for information on what \verb".bst" file to use.
The output that you get will differ slightly from that specified in the rest of this section,
but this is not a problem as long as all the relevant information is present.

\subsection{References, general}
A complete reference should provide the reader with enough information to 
locate the item concerned. Up to ten authors may be given in a particular reference; where 
there are more than ten only the first should be given followed by 
`{\it et al}'.  If you are using BibTeX
and the \verb".bst" file that you are using includes more than 10 authors, do not worry about this:
we can correct this during the production process.  Abbreviate a journal name only in accordance with the journal's
own recommendations for abbreviation---if in doubt, leave it unabbreviated.

The terms {\it loc.\ cit.}\ and {\it ibid}.\ should not be used. 
Unpublished conferences and reports should generally not be included 
in the reference list if a published version of the work exists. Articles in the course of publication should 
include the article title and the journal of publication, if known. 
A reference to a thesis submitted for a higher degree may be included 
if it has not been superseded by a published 
paper---please state the institution where the work was submitted.

The basic structure of a reference in the reference list is the same in both the alphabetical and numerical systems, the only difference the code at the start of the reference. Alphabetical references are preceded by \verb"\item[]", numerical by \verb"\bibitem{label}" or just \verb"\item" to generate a number or \verb"\nonum" where a reference is not the first in a group of references under the same number.

Note that footnotes to the text should not be 
included in the reference list, but should appear at the bottom of the relevant page by using the \verb"\footnote" command.
   
\subsection{References to journal articles}
The following guidance applies if you are producing your reference list `by hand'; that is,
without the help of BibTeX.  See section~\ref{bibby} for BibTeX help.

Article references in published articles in our journals contain three changes of 
font:
the authors and date appear in Roman type, the journal title in 
italic, the volume number in bold and the page numbers in Roman again. 
A typical journal entry would be:

\smallskip
\begin{harvard}
\item[] Spicer P E, Nijhoff F W and van der Kamp P H 2011 {\it Nonlinearity} {\bf 24} 2229
\end{harvard}
\smallskip

\noindent which would be obtained by typing, within the references
environment 
\small\begin{verbatim}
\item[] Spicer P E, Nijhoff F W and van der Kamp P H 2011 {\it Nonlinearity} 
{\bf 24} 2229
\end{verbatim}\normalsize

\noindent Features to note are the following.

\begin{enumerate}
\item The authors should be in the form of surname (with only the first 
letter capitalized) followed by the initials with no 
periods after the initials. Authors should be separated by a comma 
except for the last two which should be separated by `and' with no 
comma preceding it.

\item The year of publication follows the authors and is not in parentheses.  

\item Titles of journal articles can also be included (in Roman (upright) text after the year). Article titles are required in reference lists for {\it Inverse Problems, Journal of Neural Engineering, Measurement Science and Technology, Physical Biology, Physics in Medicine and Biology\/} and {\it Physiological Measurement}.

\item The journal is in italic and is abbreviated. If a journal has several parts denoted by 
different letters the part letter should be inserted after the journal in Roman type (e.g.\ 
{\it Phys.\ Rev.\ \rm A}). \verb"iopart.cls" includes macros for abbreviated titles of all journals handled by IOP Publishing (see table~\ref{jlab2}) and some other common titles (table \ref{jlab3}). 

\item The volume number is bold; the page number is Roman.
 Both the initial and final page numbers should be given where possible---note that for {\it Reports on Progress in Physics\/}, {\it Physiological Measurement} and {\it Physics in Medicine and Biology}
 the final page number is {\it required}. The final page number should be in 
the shortest possible form and separated from the initial page number by an en rule (\verb"--"), e.g.\ 1203--14.

\item Where there are two or more references with identical authors, 
the authors' names should be repeated for the second and subsequent references. Each individual publication should be presented as a separate reference, although in the numerical system one number can be used for several references. This facilitates linking in the online journal. 
\end{enumerate}

\subsubsection{Article numbering.}
Many journals now use article-numbering systems that do not fit the conventional {\it year-journal-volume-page numbers} pattern. Some examples are:

\numrefs{1}
\item Carlip S and Vera R 1998 {\it Phys. Rev.} D {\bf 58} 011345 
\item Davies K and Brown G 1997 {\it J. High Energy Phys.} JHEP12(1997)002
\item Hannestad S 2005 {\it J. Cosmol. Astropart. Phys.} JCAP02(2005)011
\item Hilhorst H J 2005 {\it J. Stat. Mech.} L02003
\item Gundlach C 1999 {\it Liv. Rev. Rel.} 1994-4
\endnumrefs

\noindent The website of the journal you are citing should state the correct format for citations.

\subsection{Preprint references}
Preprints may be referenced but if the article concerned has been published in a peer-reviewed journal, that reference should take precedence. If only a preprint reference can be given, it is helpful to include the article title. Examples are:
\vskip6pt
\numrefs{1}
\item Neilson D and Choptuik M 2000 {\it Class. Quantum Grav.} {\bf 17} 761 (arXiv:gr-qc/9812053)
\item Sundu H, Azizi K, S\"ung\"u J Y and Yinelek N 2013 Properties of $D_{s2}^*(2573)$ charmed-strange tensor meson arXiv:1307.6058
\endnumrefs

\noindent For preprints added to arXiv.org after April 2007 it is not necessary to include the subject area, however this information can be included in square brackets after the number if desired, e.g.
\numrefs{1}
\item Sundu H, Azizi K, S\"ung\"u J Y and Yinelek N 2013 Properties of $D_{s2}^*(2573)$ charmed-strange tensor meson arXiv:1307.6058 [hep-ph]
\endnumrefs

\subsection{References to books, conference proceedings and reports}

References to books, proceedings and reports are similar, but have only two
changes of font. The authors and date of publication are in Roman, the 
title of the book is in italic, and the editors, publisher, 
town of publication 
and page number are in Roman. A typical reference to a book and a
conference paper might be

\smallskip
\begin{harvard}
\item[] Dorman L I 1975 {\it Variations of Galactic Cosmic Rays} 
(Moscow: Moscow State University Press) p~103
\item[] Caplar R and Kulisic P 1973 {\it Proc.\
Int.\ Conf.\ on Nuclear Physics (Munich)} vol~1 (Amsterdam:  
North-Holland/American Elsevier) p~517
\end{harvard}
\smallskip

\noindent which would be obtained by with the code
\small\begin{verbatim}
\item[] Dorman L I 1975 {\it Variations of Galactic Cosmic Rays} 
(Moscow: Moscow State University Press) p~103
\item[] Caplar R and Kulisic P 1973 {\it Proc. Int. Conf. on Nuclear 
Physics (Munich)} vol~1 (Amsterdam: North-Holland/American 
Elsevier) p~517
\end{verbatim}\normalsize
\noindent 

\noindent Features to note are the following.
\begin{enumerate}
\item Book titles are in italic and should be spelt out in full with 
initial capital letters for all except minor words. Words such as 
Proceedings, Symposium, International, Conference, Second, etc should 
be abbreviated to Proc., Symp., Int., Conf., 2nd, 
respectively, but the rest of the title should be given in full, 
followed by the date of the conference and the 
town or city where the conference was held. For 
laboratory reports the laboratory should be spelt out wherever 
possible, e.g.\ {\it Argonne National Laboratory Report}.

\item The volume number, for example, vol~2, should be followed by 
the editors, if any, in the form ed~A~J~Smith and P~R~Jones. Use 
\etal if there are more than two editors. Next comes the town of 
publication and publisher, within brackets and separated by a colon, 
and finally the page numbers preceded by p if only one number is given 
or pp if both the initial and final numbers are given.

\item If a book is part of a series (for examples, {\it Springer Tracts in Modern Physics\/}), the series title and volume number is given in parentheses after the book title. Whereas for an individual volume in a multivolume set, the set title is given first, then the volume title. 
\end{enumerate}
\smallskip
\begin{harvard}
\item[]Morse M 1996 Supersonic beam sources {\it Atomic Molecular and Optical Physics\/} ({\it Experimental Methods in the Physical Sciences\/} vol 29) ed F B Dunning and R Hulet (San Diego, CA: Academic)  
\item[]Fulco C E, Liverman C T and Sox H C (eds) 2000 {\it Gulf War and Health\/} vol 1 {\it Depleted Uranium, Pyridostigmine Bromide, Sarin, and Vaccines\/} (Washington, DC: The National Academies Press)
\end{harvard}

\section{Cross-referencing\label{xrefs}}
The facility to cross reference items in the text is very useful when 
composing articles as the precise form of the article may be uncertain at the start 
and  revisions and amendments may subsequently be made. 
\LaTeX\ provides excellent facilities for doing cross-referencing
and these can be very useful in preparing articles.

\subsection{References}
\label{refs}
Cross referencing is useful for numeric reference lists because, if it 
is used, adding 
another reference to the list does not then involve renumbering all 
subsequent references. It is not necessary for referencing 
in the Harvard system where the final reference list is alphabetical 
and normally no other changes are necessary when a reference is added or
deleted.
When using \LaTeX , two passes (under certain circumstances, three passes)
are necessary initially to get the cross references right 
but once they are correct a single run is usually sufficient provided an 
\verb".aux" file is available and the file 
is run to the end each time.
If the 
reference list contains an entry \verb"\bibitem{label}", 
this command 
will produce the correct number in the reference list and 
\verb"\cite{label}" will produce the number within square brackets in the 
text. \verb"label" may contain letters, numbers 
or punctuation characters but must not contain spaces or commas. It is also
recommended that the underscore character \_{} is not used in cross
referencing. 
Thus labels of the form 
\verb"eq:partial", \verb"fig:run1", \verb"eq:dy'", 
etc, may be used. When several 
references occur together in the text \verb"\cite" may be used with 
multiple labels with commas but no spaces separating them; 
the output will be the 
numbers within a single pair of square brackets with a comma and a 
thin space separating the numbers. Thus \verb"\cite{label1,label2,label4}"
would give [1,\,2,\,4]. Note that no attempt is made by the style file to sort the 
labels and no shortening of groups of consecutive numbers is done.
Authors should therefore either try to use multiple labels in the correct 
order, or use a package such as \verb"cite.sty" that reorders labels
correctly.

The numbers for the cross referencing are generated in the order the 
references appear in the reference list, so that if the entries in the 
list are not in the order in which the references appear in the text 
then the 
numbering within the text will not be sequential. To correct this 
change the ordering of the entries in the reference list and then 
rerun the \LaTeX\ file {\it twice}.  Please ensure that all references resolve correctly: check the \verb".log" file
for undefined or multiply-defined citations, and check that the output does not contain question
marks that indicate unresolved references.

\subsection{Equation numbers, sections, subsections, figures and 
tables}
Labels for equation numbers, sections, subsections, figures and tables 
are all defined with the \verb"\label{label}" command and cross references 
to them are made with the \verb"\ref{label}" command. 

Any section, subsection, subsubsection, appendix or subappendix 
command defines a section type label, e.g. 1, 2.2, A2, A1.2 depending 
on context. A typical article might have in the code of its introduction 
`The results are discussed in section\verb"~\ref{disc}".' and
the heading for the discussion section would be:
\small\begin{verbatim}
\section{Results}\label{disc}
\end{verbatim}\normalsize
Labels to sections, etc, may occur anywhere within that section except
within another numbered environment. 
Within a maths environment labels can be used to tag equations which are 
referred to within the text. 

In addition to the standard \verb"\ref{<label>}", in \verb"iopart.cls" the abbreviated
forms given in \tref{abrefs}
are available for reference to standard parts of the text.

\Table{\label{abrefs}Alternatives to the normal references command {\tt $\backslash$ref} available in {\tt iopart.cls},
and the text generated by
them. Note it is not normally necessary to include the word equation
before an equation number except where the number starts a sentence. The
versions producing an initial capital should only be used at the start of
sentences.} 
\br
Reference&Text produced\\
\mr
\verb"\eref{<label>}"&(\verb"<num>")\\
\verb"\Eref{<label>}"&Equation (\verb"<num>")\\
\verb"\fref{<label>}"&figure \verb"<num>"\\
\verb"\Fref{<label>}"&Figure \verb"<num>"\\
\verb"\sref{<label>}"&section \verb"<num>"\\
\verb"\Sref{<label>}"&Section \verb"<num>"\\
\verb"\tref{<label>}"&table \verb"<num>"\\
\verb"\Tref{<label>}"&Table \verb"<num>"\\
\br
\endTable

\section{Tables and table captions}
Tables are numbered serially and referred to in the text 
by number (table 1, etc, {\bf not} tab. 1). Each table should have an 
explanatory caption which should be as concise as possible. If a table 
is divided into parts these should be labelled \pt(a), \pt(b), 
\pt(c), etc but there should be only one caption for the whole 
table, not separate ones for each part.

In the preprint style the tables may be included in the text 
or listed separately after the reference list 
starting on a new page. 

\subsection{The basic table format}
The standard form for a table in \verb"iopart.cls" is:
\small\begin{verbatim}
\begin{table}
\caption{\label{label}Table caption.}
\begin{indented}
\item[]\begin{tabular}{@{}llll}
\br
Head 1&Head 2&Head 3&Head 4\\
\mr
1.1&1.2&1.3&1.4\\
2.1&2.2&2.3&2.4\\
\br
\end{tabular}
\end{indented}
\end{table}
\end{verbatim}\normalsize

\noindent Points to note are:
\begin{enumerate}
\item The caption comes before the table. It should have a period at
the end.

\item Tables are normally set in a smaller type than the text.
The normal style is for tables to be indented. This is accomplished
by using \verb"\begin{indented}" \dots\ \verb"\end{indented}"
and putting \verb"\item[]" before the start of the tabular environment.
Omit these
commands for any tables which will not fit on the page when indented.

\item The default is for columns to be aligned left and 
adding \verb"@{}" omits the extra space before the first column.

\item Tables have only horizontal rules and no vertical ones. The rules at
the top and bottom are thicker than internal rules and are set with
\verb"\br" (bold rule). 
The rule separating the headings from the entries is set with
\verb"\mr" (medium rule).  These are special \verb"iopart.cls" commands.

\item Numbers in columns should be aligned on the decimal point;
to help do this a control sequence \verb"\lineup" has been defined
in \verb"iopart.cls"
which sets \verb"\0" equal to a space the size of a digit, \verb"\m"
to be a space the width of a minus sign, and \verb"\-" to be a left
overlapping minus sign. \verb"\-" is for use in text mode while the other
two commands may be used in maths or text.
(\verb"\lineup" should only be used within a table
environment after the caption so that \verb"\-" has its normal meaning
elsewhere.) See table~\ref{tabone} for an example of a table where
\verb"\lineup" has been used.
\end{enumerate}

\begin{table}
\caption{\label{tabone}A simple example produced using the standard table commands 
and $\backslash${\tt lineup} to assist in aligning columns on the 
decimal point. The width of the 
table and rules is set automatically by the 
preamble.} 

\begin{indented}
\lineup
\item[]\begin{tabular}{@{}*{7}{l}}
\br                              
$\0\0A$&$B$&$C$&\m$D$&\m$E$&$F$&$\0G$\cr 
\mr
\0\023.5&60  &0.53&$-20.2$&$-0.22$ &\01.7&\014.5\cr
\0\039.7&\-60&0.74&$-51.9$&$-0.208$&47.2 &146\cr 
\0123.7 &\00 &0.75&$-57.2$&\m---   &---  &---\cr 
3241.56 &60  &0.60&$-48.1$&$-0.29$ &41   &\015\cr 
\br
\end{tabular}
\end{indented}
\end{table}

\subsection{Simplified coding and extra features for tables}
The basic coding format can be simplified using extra commands provided in
the \verb"iopart" class file. The commands up to and including 
the start of the tabular environment
can be replaced by
\small\begin{verbatim}
\Table{\label{label}Table caption}
\end{verbatim}\normalsize
and this also activates the definitions within \verb"\lineup".
The final three lines can also be reduced to \verb"\endTable" or
\verb"\endtab". Similarly for a table which does not fit on the page when indented
\verb"\fulltable{\label{label}caption}" \dots\ \verb"\endfulltable"
can be used. \LaTeX\ optional positional parameters can, if desired, be added after 
\verb"\Table{\label{label}caption}" and \verb"\fulltable{\label{label}caption}".

\verb"\centre{#1}{#2}" can be used to centre a heading 
\verb"#2" over \verb"#1" 
columns and \verb"\crule{#1}" puts a rule across 
\verb"#1" columns. A negative 
space \verb"\ns" is usually useful to reduce the space between a centred 
heading and a centred rule. \verb"\ns" should occur immediately after the 
\verb"\\" of the row containing the centred heading (see code for
\tref{tabl3}). A small space can be 
inserted between rows of the table 
with \verb"\ms" and a half line space with \verb"\bs" 
(both must follow a \verb"\\" but should not have a 
\verb"\\" following them).
   
\Table{\label{tabl3}A table with headings spanning two columns and containing notes. 
To improve the 
visual effect a negative skip ($\backslash${\tt ns})
has been put in between the lines of the 
headings. Commands set-up by $\backslash${\tt lineup} are used to aid 
alignment in columns. $\backslash${\tt lineup} is defined within
the $\backslash${\tt Table} definition.}
\br
&&&\centre{2}{Separation energies}\\
\ns
&Thickness&&\crule{2}\\
Nucleus&(mg\,cm$^{-2}$)&Composition&$\gamma$, n (MeV)&$\gamma$, 2n (MeV)\\
\mr
$^{181}$Ta&$19.3\0\pm 0.1^{\rm a}$&Natural&7.6&14.2\\
$^{208}$Pb&$\03.8\0\pm 0.8^{\rm b}$&99\%\ enriched&7.4&14.1\\
$^{209}$Bi&$\02.86\pm 0.01^{\rm b}$&Natural&7.5&14.4\\
\br
\end{tabular}
\item[] $^{\rm a}$ Self-supporting.
\item[] $^{\rm b}$ Deposited over Al backing.
\end{indented}
\end{table}

Units should not normally be given within the body of a table but 
given in brackets following the column heading; however, they can be 
included in the caption for long column headings or complicated units. 
Where possible tables should not be broken over pages. 
If a table has related notes these should appear directly below the table
rather than at the bottom of the page. Notes can be designated with
footnote symbols (preferable when there are only a few notes) or
superscripted small roman letters. The notes are set to the same width as
the table and in normal tables follow after \verb"\end{tabular}", each
note preceded by \verb"\item[]". For a full width table \verb"\noindent"
should precede the note rather than \verb"\item[]". To simplify the coding 
\verb"\tabnotes" can, if desired, replace \verb"\end{tabular}" and 
\verb"\endtabnotes" replaces
\verb"\end{indented}\end{table}".

If all the tables are grouped at the end of a document
the command \verb"\Tables" is used to start a new page and 
set a heading `Tables and table captions'. If the tables follow an appendix then add the command \verb"\noappendix" to revert to normal style numbering.
  
\section{Figures and figure captions}

Figures (with their captions) can be incorporated into the text at the appropriate position or grouped together
at the end of the article. If the figures are at the end of the article and follow an appendix then in \verb"iopart.cls" you can add the command \verb"\noappendix" to revert to normal style numbering.  We remind you that you must seek permission
to reuse any previously-published figures, and acknowledge their use correctly---see section \ref{copyright}.

\subsection{Inclusion of graphics files\label{figinc}}
Using the \verb"graphicx" package graphics files can 
be included within figure and center environments at an 
appropriate point within the text using code such as:
\small\begin{verbatim}
\includegraphics{file.eps}
\end{verbatim}\normalsize
The \verb"graphicx" package supports various optional arguments
to control the appearance of the figure. Other similar 
packages can also be used (e.g. \verb"graphics", \verb"epsf").   Whatever package you use,
you must include it explicitly after the \verb"\documentclass" declaration using (say)
\verb"\usepackage{graphicx}".

For more detail about graphics inclusion see the documentation 
of the \verb"graphicx" package, refer to one of the books on \LaTeX , e.g. Goosens M, Rahtz S and Mittelbach F 1997 {\it The }\LaTeX\ {\it Graphics Companion\/} 
(Reading, MA: Addison-Wesley),
or download some of the excellent free documentation available via the Comprehensive
TeX Archive Network (CTAN) \verb"http://www.ctan.org"---{in particular see Reckdahl K 2006 {\it Using Imported Graphics in }\LaTeX\ {\it and }pdf\LaTeX\ \verb"http://www.ctan.org/tex-archive/info/epslatex".

IOP Publishing's graphics guidelines (available via the `Author Guidelines' link at \verb"http://authors.iop.org") provide further information on preparing \verb".eps" files.

We prefer you to use \verb".eps" files for your graphics, but we realise that converting other
formats of graphics to \verb".eps" format can be troublesome.  If you use PDF or bitmap-format graphics
such as JPG or PNG that need to be included using the pdf\TeX\ package, this is OK, but please bear
in mind that the PDF you submit should use PDF standard 1.4 or lower (use \verb"\pdfminorversion=4" at the
start of the file).

The main \LaTeX\ file must read in graphics files and subsidiary \LaTeX\ files from the current directory,
{\it not} from a subdirectory.  Your submission files are stored on our systems in a single location and we will not be able to process your
TeX file automatically if it relies on organization of the files into subdirectories.

\subsection{Captions}
Below each figure should be a brief caption describing it and, if 
necessary, interpreting the various lines and symbols on the figure. 
As much lettering as possible should be removed from the figure itself 
and included in the caption. If a figure has parts, these should be 
labelled ($a$), ($b$), ($c$), etc and all parts should be described 
within a single caption. \Tref{blobs} gives the definitions for describing 
symbols and lines often used within figure captions (more symbols are 
available when using the optional packages loading the AMS extension fonts).

\subsection{Supplementary Data}
All of our journals encourage authors to submit supplementary data attachments to 
enhance the online versions of published research articles. Supplementary data 
enhancements typically consist of video clips, animations or
data files, tables of extra information or extra figures. They can 
add to the reader's understanding and present results in attractive ways that go 
beyond what can be presented in the PDF version of the article. 
Guidelines on supplementary data file formats 
are available via the `Author Guidelines' link at \verb"http://authors.iop.org".

Software, in the form of input scripts for mathematical packages (such as Mathematica notebook files), or
source code that can be interpreted or compiled (such as Python scripts or Fortran or C programs), or executable
files, can sometimes be accepted as supplementary data, but the journal may ask you for assurances about
the software and distribute them from the article web page only subject to a disclaimer.  Contact the journal
in the first instance if you want to submit software.

\begin{table}[t]
\caption{\label{blobs}Control sequences to describe lines and symbols in figure 
captions.}
\begin{indented}
\item[]\begin{tabular}{@{}lllll}
\br
Control sequence&Output&&Control sequence&Output\\
\mr
\verb"\dotted"&\dotted        &&\verb"\opencircle"&\opencircle\\
\verb"\dashed"&\dashed        &&\verb"\opentriangle"&\opentriangle\\
\verb"\broken"&\broken&&\verb"\opentriangledown"&\opentriangledown\\
\verb"\longbroken"&\longbroken&&\verb"\fullsquare"&\fullsquare\\
\verb"\chain"&\chain          &&\verb"\opensquare"&\opensquare\\
\verb"\dashddot"&\dashddot    &&\verb"\fullcircle"&\fullcircle\\
\verb"\full"&\full            &&\verb"\opendiamond"&\opendiamond\\
\br
\end{tabular}
\end{indented}
\end{table}

\clearpage

\appendix

\section{List of macros for formatting text, figures and tables}

\begin{table}[hb]
\caption{Macros available for use in text in {\tt iopart.cls}. Parameters in square brackets are optional.}
\footnotesize\rm
\begin{tabular}{@{}*{7}{l}}
\br
Macro name&Purpose\\
\mr
\verb"\title[#1]{#2}"&Title of article and short title (optional)\\
\verb"\paper[#1]{#2}"&Title of paper and short title (optional)\\
\verb"\letter{#1}"&Title of Letter to the Editor\\
\verb"\ftc{#1}"&Title of Fast Track Communication\\
\verb"\rapid[#1]{#2}"&Title of Rapid Communication and short title (optional)\\
\verb"\comment[#1]{#2}"&Title of Comment and short title (optional)\\
\verb"\topical[#1]{#2}"&Title of Topical Review and short title 
(optional)\\
\verb"\review[#1]{#2}"&Title of review article and short title (optional)\\
\verb"\note[#1]{#2}"&Title of Note and short title (optional)\\
\verb"\prelim[#1]{#2}"&Title of Preliminary Communication \& short title\\
\verb"\author{#1}"&List of all authors\\
\verb"\article[#1]{#2}{#3}"&Type and title of other articles and 
short title (optional)\\
\verb"\address{#1}"&Address of author\\
\verb"\pacs{#1}"&PACS classification codes\\
\verb"\pacno{#1}"&Single PACS classification code\\
\verb"\ams{#1}"&Mathematics Classification Scheme\\
\verb"\submitto{#1}"&`Submitted to' message\\
\verb"\maketitle"&Creates title page\\
\verb"\begin{abstract}"&Start of abstract\\
\verb"\end{abstract}"&End of abstract\\
\verb"\nosections"&Inserts space before text when no sections\\
\verb"\section{#1}"&Section heading\\
\verb"\subsection{#1}"&Subsection heading\\
\verb"\subsubsection{#1}"&Subsubsection heading\\
\verb"\appendix"&Start of appendixes\\
\verb"\ack"&Acknowledgments heading\\
\verb"\References"&Heading for reference list\\
\verb"\begin{harvard}"&Start of alphabetic reference list\\
\verb"\end{harvard}"&End of alphabetic reference list\\
\verb"\begin{thebibliography}{#1}"&Start of numeric reference list\\
\verb"\end{thebibliography}"&End of numeric reference list\\
\verb"\etal"&\etal for text and reference lists\\
\verb"\nonum"&Unnumbered entry in numerical reference list\\
\br
\end{tabular}
\end{table}

\clearpage

\begin{table}
\caption{Macros defined within {\tt iopart.cls}
for use with figures and tables.}
\begin{indented}
\item[]\begin{tabular}{@{}l*{15}{l}}
\br
Macro name&Purpose\\
\mr
\verb"\Figures"&Heading for list of figure captions\\
\verb"\Figure{#1}"&Figure caption\\
\verb"\Tables"&Heading for tables and table captions\\
\verb"\Table{#1}"&Table caption\\
\verb"\fulltable{#1}"&Table caption for full width table\\
\verb"\endTable"&End of table created with \verb"\Table"\\
\verb"\endfulltab"&End of table created with \verb"\fulltable"\\
\verb"\endtab"&End of table\\
\verb"\br"&Bold rule for tables\\
\verb"\mr"&Medium rule for tables\\
\verb"\ns"&Small negative space for use in table\\
\verb"\centre{#1}{#2}"&Centre heading over columns\\
\verb"\crule{#1}"&Centre rule over columns\\
\verb"\lineup"&Set macros for alignment in columns\\
\verb"\m"&Space equal to width of minus sign\\
\verb"\-"&Left overhanging minus sign\\
\verb"\0"&Space equal to width of a digit\\
\br
\end{tabular}
\end{indented}
\end{table}

\clearpage

\begin{table}[hb]
\caption{\label{jlab2}Abbreviations in {\tt iopart.cls} for journals handled by IOP Publishing.}
\footnotesize
\begin{tabular}{@{}llll}
\br
{\rm Short form of journal title} & Macro & {\rm Short form of journal title} & Macro \\
\mr
2D Mater.&\verb"\TDM"&J. Radiol. Prot.&\verb"\JRP"\\
AJ&\verb"\AJ"&J. Semicond.&\verb"\JOS"\\
ApJ&\verb"\APJ"&J. Stat. Mech.&\verb"\JSTAT"\\
ApJL&\verb"\APJL"&Laser Phys.&\verb"\LP"\\
ApJS&\verb"\APJS"&Laser Phys. Lett.&\verb"\LPL"\\
Adv. Nat. Sci: Nanosci. Nanotechnol.&\verb"\ANSN"&Metrologia&\verb"\MET"\\
Appl. Phys. Express&\verb"\APEX"&Mater. Res. Express&\verb"\MRE"\\
Biofabrication&\verb"\BF"&Meas. Sci. Technol.&\verb"\MST"\\
Bioinspir. Biomim.&\verb"\BB"&Methods Appl. Fluoresc.&\verb"\MAF"\\
Biomed. Mater.&\verb"\BMM"&Modelling Simul. Mater. Sci. Eng.&\verb"\MSMSE"\\
Chin. J. Chem. Phys.&\verb"\CJCP"&Nucl. Fusion&\verb"\NF"\\
Chinese Phys. B&\verb"\CPB"&New J. Phys.&\verb"\NJP"\\
Chinese Phys. C&\verb"\CPC"&Nonlinearity&\verb"\NL"\\
Chinese Phys. Lett.&\verb"\CPL"&Nanotechnology&\verb"\NT"\\
Class. Quantum Grav.&\verb"\CQG"&Phys. Biol.&\verb"\PB"\\
Commun. Theor. Phys.&\verb"\CTP"&Phys. Educ.&\verb"\PED"\\
Comput. Sci. Disc.&\verb"\CSD"&Phys.-Usp.&\verb"\PHU"\\
Environ. Res. Lett.&\verb"\ERL"&Physiol. Meas.&\verb"\PM"\\
EPL&\verb"\EPL"&Phys. Med. Biol.&\verb"\PMB"\\
Eur. J. Phys.&\verb"\EJP"&Phys. Scr.&\verb"\PS"\\
Fluid Dyn. Res.&\verb"\FDR"&Plasma Phys. Control. Fusion&\verb"\PPCF"\\
Inverse Problems&\verb"\IP"&Plasma Sci. Technol.&\verb"\PST"\\
Izv. Math.&\verb"\IZV"&Plasma Sources Sci. Technol.&\verb"\PSST"\\
Jpn. J. Appl. Phys.&\verb"\JJAP"&Quantum Electron.&\verb"\QEL"\\
J. Breath Res.&\verb"\JBR"&Rep. Prog. Phys.&\verb"\RPP"\\
JCAP&\verb"\JCAP"&Res. Astron. Astrophys.&\verb"\RAA"\\
J. Geophys. Eng.&\verb"\JGE"&Russ. Chem. Rev.&\verb"\RCR"\\
JINST&\verb"\JINST"&Russ. Math. Surv.&\verb"\RMS"\\
J. Micromech. Microeng.&\verb"\JMM"&Sb. Math.&\verb"\MSB"\\
J. Neural Eng.&\verb"\JNE"&Science Foundation in China&\verb"\SFC"\\
J. Opt.&\verb"\JOPT"&Sci. Technol. Adv. Mater.&\verb"\STAM"\\
J. Phys. A: Math. Theor.&\verb"\jpa"&Semicond. Sci. Technol.&\verb"\SST"\\
J. Phys. B: At. Mol. Opt. Phys.&\verb"\jpb"&Smart Mater. Struct.&\verb"\SMS"\\
J. Phys: Condens. Matter&\verb"\JPCM"&Supercond. Sci. Technol.&\verb"\SUST"\\
J. Phys. D: Appl. Phys.&\verb"\JPD"&Surf. Topogr.: Metrol. Prop.&\verb"\STMP"\\
J. Phys. G: Nucl. Part. Phys.&\verb"\jpg"&Transl. Mater. Res.&\verb"\TMR"\\
\mr
{\it IOP Conference Series journals}\\
\mr
J. Phys.: Conf. Ser.&\verb"\JPCS"\\
IOP Conf. Ser.: Earth Environ. Sci.&\verb"\EES"\\
IOP Conf. Ser.: Mater. Sci. Eng.&\verb"\MSE"\\

\br
\end{tabular}

\end{table}

\clearpage

\begin{table}

\caption{\label{jlab2b}Abbreviations for IOP Publishing journals that are no longer published.}
\begin{indented}
\item[]\begin{tabular}{@{}lll}
\br
{\rm Short form of journal title} & Macro name & Years relevant\\
\mr
J. Phys. A: Math. Gen.&\verb"\JPA"&1975--2006\\
J. Phys. B: At. Mol. Phys.&\verb"\JPB"&1968--1987\\
J. Phys. C: Solid State Phys.&\verb"\JPC"&1968--1988\\
J. Phys. E: Sci. Instrum.&\verb"\JPE"&1968--1989\\
J. Phys. F: Met. Phys.&\verb"\JPF"&1971--1988\\
J. Phys. G: Nucl. Phys.&\verb"\JPG"&1975--1988\\
Pure Appl. Opt.&\verb"\PAO"&1992--1998\\
Quantum Opt.&\verb"\QO"&1989--1994\\
Quantum Semiclass. Opt.&\verb"\QSO"&1995--1998\\
J. Opt. A: Pure Appl. Opt.&\verb"\JOA"&1999--2009\\
J. Opt. B: Quantum Semiclass. Opt.&\verb"\JOB"&1999--2005\\
\br
\end{tabular}
\end{indented}
\end{table}

\begin{table}[h]

\caption{\label{jlab3}Abbreviations in {\tt iopart.cls} for some 
common journals not handled by IOP Publishing.}
\begin{indented}
\item[]\begin{tabular}{@{}llll}
\br
Short form of journal & Macro & Short form of Journal & Macro\\
\mr
Acta Crystallogr.&\verb"\AC"&J. Quant. Spectrosc. Radiat. Transfer&\verb"\JQSRT"\\
Acta Metall.&\verb"\AM"&Nuovo Cimento&\verb"\NC"\\
Ann. Phys., Lpz&\verb"\AP"&Nucl. Instrum. Methods&\verb"\NIM"\\
Ann. Phys., NY&\verb"\APNY"&Nucl. Phys.&\verb"\NP"\\
Ann. Phys., Paris&\verb"\APP"&Phys. Fluids&\verb"\PF"\\
Can. J. Phys.&\verb"\CJP"&Phys. Lett.&\verb"\PL"\\
Gen. Rel. Grav.&\verb"\GRG"&Phys. Rev.&\verb"\PR"\\
J. Appl. Phys.&\verb"\JAP"&Phys. Rev. Lett.&\verb"\PRL"\\
J. Chem. Phys.&\verb"\JCP"&Proc. R. Soc.&\verb"\PRS"\\
J. High Energy Phys.&\verb"\JHEP"&Phys. Status Solidi&\verb"\PSS"\\
J. Magn. Magn. Mater.&\verb"\JMMM"&Phil. Trans. R. Soc.&\verb"\PTRS"\\
J. Math. Phys.&\verb"\JMP"&Rev. Mod. Phys.&\verb"\RMP"\\
J. Opt. Soc. Am.&\verb"\JOSA"&Rev. Sci. Instrum.&\verb"\RSI"\\
J. Physique&\verb"\JP"&Solid State Commun.&\verb"\SSC"\\
J. Phys. Chem.&\verb"\JPhCh"&Sov. Phys.--JETP&\verb"\SPJ"\\
J. Phys. Soc. Jpn.&\verb"\JPSJ"&Z. Phys.&\verb"\ZP"\\

\br
\end{tabular}
\end{indented}
\end{table}

\clearpage

\section{Including author names using Chinese, Japanese and Korean characters in submissions to IOP Publishing journals}
Authors in all IOP Publishing journals have the option to include names in Chinese, Japanese or Korean (CJK) characters in addition to the English name. The names will be displayed in the print issue and the online PDF, abstract and table of contents, in parentheses after the English name. 

It is the decision of the individual authors whether or not to include a CJK version of their names; for a single article it is not necessary for all authors to include a CJK name if only one author wishes to do so. It is the responsibility of the authors to check the accuracy and formatting of the names in the final proofs that they receive prior to publication. 

To include names in CJK characters, authors should use the \verb"cjk.sty" package, available from \verb"http://www.ctan.org/tex-archive/language/chinese/CJK/". Users should be aware that this is a very large and complicated package which relies on a large number of fonts.  We recommend using a TeX package that includes this package and all of the fonts by default, so that manual configuration is not required (e.g. the TeXLive distribution, which is available on all platforms (Macintosh, Windows and Linux)).

The documentation for the \verb"cjk.sty" package gives information on how CJK characters can be included in TeX files.  Most authors will find it convenient to include the characters in one of the standard encodings such as UTF-8, GB or JIS, if they have access to a text editor that supports such encodings.

Example TeX coding might be:
\begin{verbatim}
\documentclass[12pt]{iopart}
\usepackage{CJK}
.
.
.
\begin{document}
\begin{CJK*}{GBK}{ }

\title[]{Title of article}
\author{Author Name (CJK characters)}
\address{Department, University, City, Country}
.
.
.
\end{CJK*}
\end{verbatim}

To avoid potential problems in handling the CJK characters in submissions, authors should always include a PDF of the full version of their papers (including all figure files, tables, references etc) with the CJK characters in it.

\fi

\end{document}